\documentclass{aa}  

\usepackage{graphicx}
\usepackage{txfonts}
\usepackage{hyperref}
\usepackage{xcolor}
\begin{document}

   \title{Near-core magnetic field strengths inferred from gravity modes in intermediate-mass stars}

   \subtitle{}

   \author{O. Dürfeldt-Pedros
          \inst{1,*}
          \and
          V. Antoci\inst{1}
          \and
          D. Lecoanet\inst{2, 3}
          \and 
          Z. Guo\inst{4}
          \and 
          J. Labadie-Bartz\inst{1,5}
          }
   \institute{DTU Space, Technical University of Denmark, Elektrovej 327, Kgs. Lyngby, 2800, Denmark \\ \email{odp@space.dtu.dk}, \email{antoci@space.dtu.dk} \and Department of Engineering Sciences and Applied Mathematics, Northwestern University, Evanston IL 60208, USA \and CIERA, Northwestern University, Evanston IL 60201, USA \and
   Department of Applied Mathematics, School of Mathematics, University of Leeds, Leeds LS2 9JT, UK \and
   LIRA, Observatoire de Paris, Universit\'e PSL, CNRS, Sorbonne Universit\'e, Universit\'e Paris Cit\'e, CY Cergy Paris Universit\'e, 92190 Meudon, France
}

   \date{Received MONTH DAY, YEAR; accepted MONTH DAY, YEAR}

 
  \abstract
   {}
   {In this work, we derive upper limits for the strength of the near-core magnetic field in intermediate-mass stars, based on the fact that high-order $g$-modes can be fully suppressed by a critical magnetic field. We consider both purely poloidal magnetic fields and mixed geometries with the inclusion of a toroidal component. We examine how the upper limits on magnetic field strengths are affected by the degree and azimuthal order of the oscillations, as well as the magnetic field configuration. We compare our results with observations of red giant stars, which are the evolved counterparts of main sequence intermediate-mass stars, to assess the consistency of the field strengths across evolutionary stages.}
   {We consider two $\gamma$ Doradus stars hosting high-order $g$-modes and an evolved $\delta$ Scuti star with mixed modes. These targets have prior mode identification from observations, which allows us to use these mode frequencies to probe the near-core region. We determine the best structural model from their stellar parameters through grid-based modeling with MESA. Frequencies for the best models are extracted using GYRE and matched to the observed modes. The critical magnetic fields for all calculated frequencies in our models are obtained from the Dedalus code, from which we can infer an upper limit on the near-core field strength.}
   {We find an upper limit on the near-core radial field strength of $B_r \approx 130\,\rm{kG}$ and $B_r \approx 13\,\rm{kG}$, assuming a dipole field configuration, for the two $\gamma$ Doradus stars KIC~3127996 and KIC~5876187, respectively. For 44 Tau, analysis of mixed modes yields a field strength of $B_r \approx 1771\,\rm{kG}$. Different magnetic field configurations and mode degrees lead to different estimates. The results for the radial component of the magnetic field in the main sequence $\gamma$ Doradus stars are consistent with estimates of magnetic field strengths in red giant stars that assume an internal field generated by a core dynamo, although the stronger of the two inferred magnetic fields may require some enhancement by a fossil field. The toroidal component does not affect $g$-modes significantly and is required to be more than 200 times stronger than the radial component in order to suppress $g$-modes. We find evidence of rotational modulation in the light curves of the two $\gamma$ Doradus stars, from which we are able to show that the rotation rate is consistent with rigid body rotation in the envelope.}
   {}


   \maketitle

\section{Introduction}\label{sec:Introduction}
Asteroseismic observations of intermediate-mass (spanning late F-type through A and B types) stars have revealed that a significant fraction of these experience near-rigid body rotation \citep{Mosser2012b, Saio2021}, which is in contrast with theoretical expectations and suggests that angular momentum transport is more efficient than expected \citep{Eggenberger2012, Marques2013, Ceillier2013}. To explain this, magnetic fields and associated instabilities are proposed as a plausible mechanism for the efficient transport of angular momentum \citep{Maeder2005, Cantiello2014}. For bright intermediate-mass stars, surface magnetic fields have been detected through spectropolarimetry, with strengths typically of 100 G to 1 kG and a simple dipole field configuration \citep{Bagnulo2006, Auriere2007}. For high-mass to intermediate-mass stars, the incidence of strong, global surface fields was found to be roughly 10\% \citep{Moss2001, Wade2014}. However, very weak fields of the order 1 to 10 G have been found \citep{Petit2011, Lignieres2014, Neiner2015, Blazere2016, Neiner2017}, and other proxies for magnetic fields have helped identify intermediate-mass stars hosting these without direct detections \citep{Antoci2025}, suggesting that they may be more widespread in intermediate-mass stars.

Although magnetic fields are considered one of the dominant mechanisms for increased angular momentum transport, their nature, structure, and strength remain uncertain \citep[see e.g.][]{Braithwaite2017}. To explain the observed surface fields, several mechanisms are considered. Firstly, observations could suggest that stars host strong global fields of fossil origin, which could be stable over the timescale of main sequence evolution \citep{Braithwaite2004}. However, fossil fields are only found in a minority of stars and cannot by themselves explain the rotation profiles of the intermediate-mass population. Secondly, the Tayler-Spruit dynamo has been proposed to explain the generation of a magnetic field in the radiative zones of stars in the presence of differential rotation \citep{Spruit2002, Maeder2005, Fuller2019}. Thirdly, sub-surface convection is believed to be capable of generating a surface magnetic field and produce the observational features found in hump and spike stars, where the observed spike near the rotation frequency in the amplitude spectra is associated with the existence of stellar spots \citep{Cantiello2019, Henriksen2023, Antoci2025}. These are locally generated fields and therefore not large scale, as is the case with stable fossil fields. Finally, three-dimensional magnetohydrodynamic (MHD) simulations have shown that core dynamos can produce stable magnetic fields \citep{Brun2005, Augustson2016}. However, the time scales required for these to reach the surface via magnetic diffusion are predicted to  be longer than the main sequence \citep{Schuessler1978, Moss1989}, and core dynamos cannot explain the structured global fields observed at the surface of some intermediate-mass stars \citep[e.g.][]{Hidalgo2024}. MHD simulations have examined how a global dipolar fossil field in the radiative envelope can couple with a core dynamo-generated field, enhancing the latter and enforcing rigid-body rotation \citep{Featherstone2009, Hidalgo2025}. Other calculations looked into how differential rotation between the convective core and radiative envelope can lead to a strong toroidal component of the magnetic field \citep{Ratnasingam2024}. It follows from this summary that the study of magnetic fields is a highly active research area and that we need additional constraints from observations to assess our theories and models. 

Asteroseismology has emerged as a powerful tool to understand the internal structure of stars and has recently provided insight into internal magnetic fields through the analysis of gravity ($g$) and mixed modes. Such oscillations are sensitive to the near-core region and allow us to probe conditions and processes occurring deep inside stars. $g$-modes are sensitive to the chemical composition gradient left behind by the receding core in intermediate-mass stars, which translates into a periodic dip in the period spacing pattern \citep{Miglio2008}. In intermediate-mass stars, it has been shown that the period spacing pattern is altered by rotation in an almost linear manner \citep{Bouabid2013, VanReeth2015}. Analysis of these patterns has thereby allowed us to estimate the rotation frequency of the near-core region \citep[e.g.][]{VanReeth2015, VanReeth2016, VanReeth2018, Ouazzani2017, Li2020}, and in the case where $g$-modes couple with inertial modes, the rotation frequency of the core itself \citep{Ouazzani2020, Saio2021}. When combined with surface rotation frequencies (e.g. determined from spots) and envelope rotation frequencies (measured from pressure modes), a comprehensive picture of stellar rotation profiles can be painted. A similar effect on the period spacings has been found for magnetic fields, which introduce a curvature in the pattern \citep{Prat2019, Prat2020, VanBeeck2020, Dhouib2022, Rui2024}.

In red giants, it has been shown that magnetic fields above $100\,\rm{kG}$ can account for the amplitude suppression of mixed modes, where part of the mode energy is lost to a magnetic greenhouse effect \citep{Fuller2015, Stello2016, Lecoanet2017, Loi2020b}. If the magnetic field is strong enough, above a critical value that depends on the internal structure and the oscillation frequency, magnetic tension causes the gravity waves to convert to magnetic waves, which completely dissipate as they propagate through the star. We define this value as the critical magnetic field strength. \citet{Lecoanet2022} utilized the interactions between $g$-modes and magnetic fields described by \citet{Fuller2015} and \citet{Lecoanet2017} to estimate the radial component of the near-core magnetic field strength for HD 43317, a Slowly Pulsating B-star (SPB) with a directly detected globally organized, strong surface magnetic field $B_{\rm{p}} = 1312 \pm 332\,\rm{G}$. They derive an upper limit for a dipole field strength near the convective core, $B_r \approx 500\,\rm{kG}$, by comparing with the observations from \citet{Buysschaert2018} and assuming that the high-order $g$-modes are fully suppressed by a critical magnetic field. Additional studies have revealed that weaker internal magnetic fields in the cores of red giants can produce frequency shifts \citep{Bugnet2021}. These shifts lead to asymmetries in the rotationally split multiplets of a mode, which are only consistent with magnetic fields with radial components $B_r=25\rm{-}150\,\rm{kG}$ \citep{Li2022}. Core magnetic fields have also been detected in red giant stars by analyzing the deviations in the gravity mode period spacing patterns, with strengths up to 600\,kG \citep{Deheuvels2023}. Recent work suggests that core magnetic fields can also be detected from a perturbative analysis of $g$-mode frequency shifts in $\gamma$ Doradus stars \citep{Lignieres2024, Takata2026}. This led to the first detection of an internal magnetic field in a main sequence F-type star, with the radial component $B_r=3.5\pm0.1\,\rm{kG}$ and an even stronger toroidal component  \citep{Takata2026}, consistent with the 3D simulations of \citet{Ratnasingam2024}. 

In this work, we apply the method presented by \citet{Lecoanet2022} to a sample of intermediate-mass stars in the range $1.4\rm{-}2\, M_\odot$ that host $g$-mode oscillations. They are progenitors of red giants stars, which have been studied in great detail in recent years. Analyzing the main sequence counterparts is essential to assess the consistency of our theories and better understand stellar evolution \citep{Skoutnev2025}. This work focuses on inferring near-core magnetic field strengths for different types of pulsating stars in the instability strip. We consider targets for which mode identification has already been carried out based on observations. We use MESA and GYRE to generate stellar models and extract model-specific oscillation frequencies. We study two $\gamma$ Doradus stars, which are main sequence stars hosting high-order $g$-modes, as well as one evolved $\delta$ Scuti star, where mixed modes are observed. We explore critical magnetic field estimates in these stars for different mode degrees, azimuthal orders, and magnetic field configurations. We present the targets selected for this analysis in Section \ref{sec:SectionI}, before outlining the methodology for determining critical magnetic field strengths in Section \ref{sec:SectionII}. Results are shown in Section \ref{sec:Results} and discussed in Section \ref{sec:Discussion}, before concluding in Section \ref{sec:Conclusions}.  

\begin{table*}
    \centering
    \caption{Summary table of the targets studied in this work.}
    \begin{tabular}{ccccccccccc}
        \hline 
        Target ID & Variability & $T_{\rm{eff}}$ [K] & $L_\odot$ & $f_{\rm{rot}}\ [\rm{d}^{-1}]$ & $\Pi_0$ [s] & M $[M_\odot]$ & R $[R_\odot]$ & $X_{\rm{c}}$ & $f_0\ [\rm{d^{-1}}]$ & $f_1\ [\rm{d^{-1}}]$ \\
        \hline
        KIC~3127996 & $\gamma$ Doradus & 7307 & 7.47 & $0.0518$ & $4007$ & - & - & - & - & - \\
        & & $\pm 300$ & $\pm 0.5$ & $\pm 0.0007$ & $\pm 2$ & & & & \\
        Best Model &  & 6727 & 6.30 & - & 4198 & 1.50 & 1.84 & 0.41 & - & - \\
        \hline
        KIC~5876187 & $\gamma$ Doradus & 6783 & 7.99 & $0.596$ & $4100$ & - & - & - & - & -\\
        & & $\pm 110$ & $\pm 0.5$ & $\pm 0.005$ & $\pm 100$ & & & & & \\
        Best Model &  & 6612 & 8.30 & - & 4257 & 1.58 & 2.19 & 0.30 & - & - \\
        \hline
        44 Tau & $\delta$ Scuti & 6900 & 20.18 & - & - & - & $3.23$ & - & 6.898 & 8.961 \\
        & & $\pm 100$ & $\pm 1.16$ & & & & $\pm 0.13$ & & & \\
        Best Model &  & 6858 & 21.64 & - & - & 1.86 & 3.29 & - & 6.917 & 8.986 \\
        \hline
    \end{tabular}
    \tablefoot{The stellar parameters, pulsation class and rotation frequency is shown for all stars. For the $\gamma$ Doradus stars, we include the asymptotic period spacing used for modeling. For 44 Tau, we instead use the radial fundamental and first overtone frequencies. The uncertainties for the rotation frequencies are taken from \citet{Li2020}. The values obtained from the best model are included, where we also give the derived mass, radius and hydrogen content in the core.}
    \label{tab:TargetSummary}
\end{table*}

\section{Target selection}\label{sec:SectionI}
We focus on targets with prior mode identification. The two $\gamma$ Doradus stars, KIC~3127996 and KIC~5876187, are taken from the sample of 611 targets observed by \textit{Kepler} \citep{Borucki2010, Koch2010} and presented in \citet{Li2020}. Data for the evolved $\delta$ Scuti star, 44 Tau, is collected from \citet{Lenz2010}. A parameter summary for each
target is provided in Table \ref{tab:TargetSummary}. No surface magnetic field detections have been reported in the literature for any of the considered stars. We discuss these targets separately below. 

\subsection{KIC 3127996}\label{sec:KIC3127996}
KIC 3127996 is a $\gamma$ Doradus star that exhibits rotational splitting of dipole $(\ell,m) = (1,-1),(1,0),(1,1)$ and quadrupole $(\ell,m) = (2,-2),(2,-1),(2,1)$ $g$-modes \citep{Li2020}.  We use the sign convention that positive (negative) values of the azimuthal order $m$ correspond to prograde (retrograde) modes and that negative values of $n$ represent $g$-modes, as is done in GYRE. A summary of the observed frequencies can be found in Table \ref{tab:KIC3217996}. The observed orders from our frequency extraction (see Section \ref{sec:FreqExtract}) $n_{\rm{pg}}$ span the ranges $[-31; -20]$ and $[-36; -28]$ for the $\ell = 1$ and $\ell = 2$ splittings, respectively. The asymptotic period spacing was found to be $\Pi_0 = 4007 \pm 2\, \rm{s}$ \citep{Li2020}. We note that the uncertainty is unrealistically low for our modeling purposes since we use this parameter to select our best models (see Section \ref{sec:SectionII}), and instead choose to introduce $\sigma_{\Pi_0} = 200$ s, which corresponds to an uncertainty of 5\%. 

It appears that the effective temperature from the \textit{Kepler} DR25 shown in \citet{Li2020} is overestimated by roughly 700~K in the case of KIC~3127996 when compared to other values from literature \citep{Frasca2016} and may be in a binary system. We use \textit{Gaia} DR3 \citep{GaiaDR3} instead, since the effective temperature agrees better with other studies, and we can obtain an estimate of the luminosity. We therefore have $T_{\rm{eff}} = 7307~\rm{K}$ and $L = 7.47 L_\odot$. The uncertainties on the luminosity and temperature are set to $\sigma_{L_\odot} = 0.5\ L_\odot$ \citep{Fouesneau2023} and $\sigma_{T_{\rm{eff}}} = 300$~K. 

\subsection{KIC 5876187}\label{sec:KIC5876187}
KIC 5876187 hosts $(\ell, m) = $ (1, 1), (2, 2), (3, 3) sectoral pulsation modes. The observed frequencies and mode identification are provided in Table \ref{tab:KIC5876187}, with observed orders spanning the range $[-111; -55]$ based on our frequency extraction. We choose this target to test how the critical magnetic field strength estimations deviate from one degree to another and whether it is possible to infer consistent values for all modes. We use the \textit{Gaia} DR3 values $T_{\rm{eff}} = 6783 \rm{K}$ and $L = 7.99 L_\odot$ and note that the effective temperature is consistent with \citet{Li2020} within the recommended uncertainties for \textit{Gaia} DR3 \citep{Fouesneau2023}.

\subsection{44 Tau}\label{sec:44Tau}
44 Tau is a bright star hosting $\delta$ Scuti pulsations. The stellar parameters are $T_{\rm{eff}} = 6900 \pm 100 \rm{K}$ and $L = 20.18 \pm 1.16 L_\odot$ \citep{Lenz2010}. These values were obtained from ground-based observations. Detailed mode identification and modeling reveal that 44 Tau is most likely an evolved star located in the overall contraction phase in the HR diagram, though several models with masses in the range $1.8M_\odot$ to $2 M_\odot$ are able to reproduce the observed frequencies \citep{Lenz2010}. \citet{Antoci2007} and \citet{Zima2007} show that 44 Tau has low values for $v\sin i$. Although this only provides a lower limit on the rotation frequency of the star when the inclination is unknown, we consider that 44 Tau most likely is a slow rotator and model it as such.

Mixed modes have been identified in the amplitude spectrum for this star, allowing us to probe the near-core region. We provide in \autoref{tab:44TauSummary} the list of identified frequencies from \citet{Lenz2010} that are used in our modeling. Since these behave as $g$-modes in the deep interior of the star, we test their ability to inform us on the critical magnetic field strength. This allows us to examine the requirements on the magnetic field strengths to fully suppress pure $g$-modes in $\delta$ Scuti stars and whether this could explain the differences between $\delta$ Scuti and hybrid pulsators, which host both $g$-modes and p-modes. 

\section{Methods}\label{sec:SectionII}
In this section, we present the methods used to obtain the results presented in this paper. We highlight here that the aim is not to perform detailed forward modeling of the selected stars, but to find a reliable representation of each target with a limited number of free parameters and infer critical magnetic field strengths assuming different field configurations. For this reason, we limit the number of degrees of freedom in our models. Additional results coming from the frequency analysis of observational data are also provided, since these offer deeper insight into our targets. 

We model our targets assuming rigid-body rotation. This is motivated by observational evidence from $g$-modes, showing that a significant number of $\gamma$ Doradus stars have uniformly rotating envelopes \citep{Li2020}, and inertial modes, which highlight the same behavior in the core \citep{Saio2021}. For the two $\gamma$ Doradus stars considered in this work, we constrain the surface rotation rate from evidence of rotational modulation in the \textit{Kepler} light curves, which further supports this decision (see Section \ref{sec:Modulation}). We therefore use the near-core rotation frequency derived by \citet{Li2020} from the slope of the period spacing patterns as the global rotation frequency $f_{\rm{rot}}$ of the star in our models. In the case of 44 Tau, the surface rotation is known, but we are unable to determine the near-core rotation rate and simply assume it to be uniform.

\subsection{\textit{Kepler} frequency extraction}\label{sec:FreqExtract}
\begin{figure*}
    \centering
    \begin{minipage}{0.49\textwidth}
        \centering
        \includegraphics[width=\linewidth]{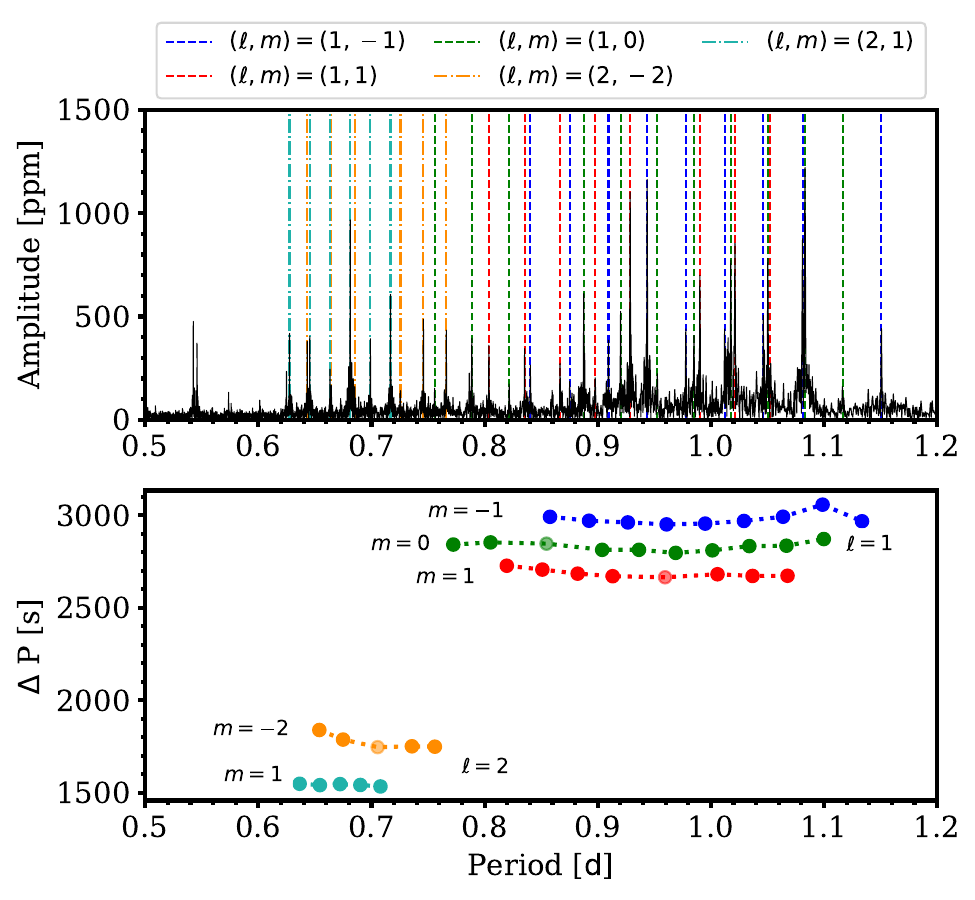}
    \end{minipage}
    \hfill
    \begin{minipage}{0.49\textwidth}
        \centering
        \includegraphics[width=\linewidth]{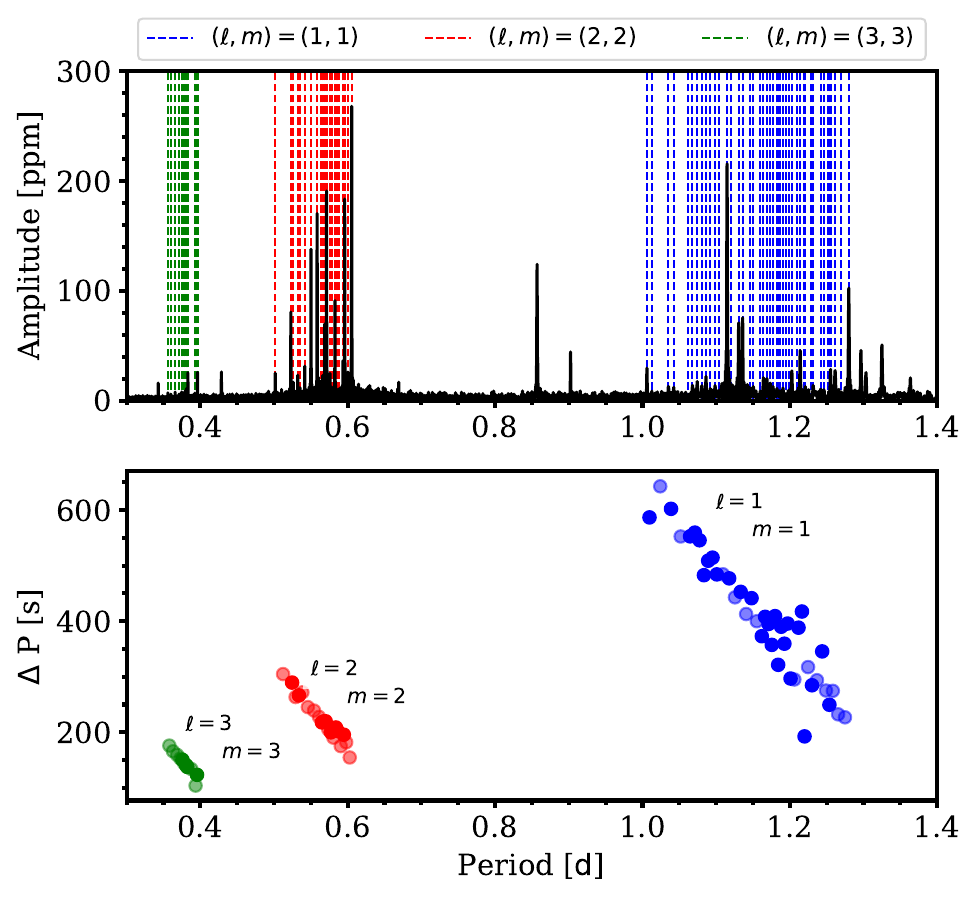}
    \end{minipage}
    \caption{\textit{Upper:} Amplitude spectra for the $\gamma$ Doradus stars KIC~3127996 (left) and KIC~5876187 (right). The frequencies extracted according to the method described in Section \ref{sec:FreqExtract} are highlighted with dashed lines and color-coded according to the mode identification. \textit{Lower:} Period spacing patterns for both stars. Full circles represent the period spacings that were used to compute the asymptotic period spacing. Fainter points highlight the missing frequencies to complete the observed pattern where the period and $\Delta$ P values are taken as the average between the two adjacent detected frequencies according to how many consecutive signals are missing.}
    \label{fig:KIC5871687_ampspec}
\end{figure*}

While the range of radial orders for a given combination of $l$ and $m$ are provided in \citet{Li2020} for KIC~3127996 and KIC~5876187, the observed frequencies and their associated $S/N$ ratios are not. The \textit{Kepler} light curves were downloaded from the Mikulski Archive for Space Telescopes (MAST\footnote{\url{https://archive.stsci.edu/}}). After removing outliers, normalizing each segment, and stitching together the light curves, the data were iteratively pre-whitened using the \textsc{Astropy} \citep{astropy2013,astropy2018} LombScargle package to select all frequencies down to a threshold of $S/N = 4$. 
The noise level was calculated as the median value within the frequency range of interest at each iteration. For KIC 5876187 this range was between 0.3 and 1.4 d and for KIC 3127996 between 0.5 and 1.2 d (i.e. the period range plotted in Fig.~\ref{fig:KIC5871687_ampspec}). The specific choice of the $S/N$ threshold has no appreciable effect on the results. For KIC 3127996 all signals involved in the period spacing patterns have $S/N > 15$ and there are no marginal signals ($S/N$ between $\sim$4 -- 10) that seem to belong to any of the identified period spacing patterns. For KIC 5876187 the first and last signals in a given group have $S/N > 20$ except for the shortest period signal with $S/N \approx 5.5$. Only seven peaks (out of a total of 75) identified in the period spacing patterns for KIC 5876187 have $S/N$ between 4 and 5. 

The extracted signals were then analyzed with the \texttt{FLOSSY}\footnote{\url{https://github.com/IvS-KULeuven/FLOSSY}} package to manually identify period spacing patterns with common $l$ and $m$ values following the approach of \citet{Garcia2022} and guided by the results published in \citet{Li2020}. To further verify the frequencies that were extracted and their mode identification, each set of frequencies with the same mode geometry was analyzed using the \texttt{morse}\footnote{\url{https://github.com/schristophe/morse}} Python package \citep{Christophe2018}, from which we determined $\Pi_0$. For both cases, our values of $\Pi_0$ are within 5\% of the values reported by \citet{Li2020}. We show the amplitude spectra obtained from the \textit{Kepler} data along with the extracted frequencies in the upper panels of Fig. \ref{fig:KIC5871687_ampspec}. The period spacing patterns are provided in the lower panels for guidance. When modeling our targets, we determined the closest matching frequency from our models with the observed modes for each $(\ell,m)$ pair. We refer to Section \ref{sec:GYRE} for further details. For KIC~3127996 we do not detect the $(\ell,m)=(2,-1)$ family since the frequencies shown in \citet{Li2020} are not statistically significant for our $S/N$ criterion. All frequencies used in this work are provided in \autoref{sec:Frequencies}.

\subsection{MESA grid}\label{sec:MESAGrid}
We use the stellar evolutionary code MESA \citep{Paxton2011, Paxton2013, Paxton2015, Paxton2018, Paxton2019, Jermyn2023}, release r24.08.01 \citep{Paxton2024}, to compute models for a range of stellar masses, creating a grid of models from which we need to select the best-suited one for each star in our sample. Our MESA inlist template is provided in \autoref{sec:MESAinlist} and can also be found on the MESA marketplace \footnote{\url{https://mesastar.org/marketplace/inlists}}. The evolutionary tracks cover the mass range $1.4$ to $2.0\ M_\odot$ in steps of $0.01\ M_\odot$. The MESA models considered do not include rotation, thus we consider solely spherically symmetric stars. We use the \texttt{pp\_cno\_extras\_o18\_ne22} reaction net to compute the nuclear reactions from the pp-chain and CNO cycles at each time step. The OPAL opacity tables \citep{Iglesias1993} are used to compute the opacities in our models. The chemical composition is set with the GS98 solar metal mixture reference values \citep{GS98}, where we consider only solar metallicity ($Z_{\rm{init}} = 0.02$) for our grid. We do not include the effect of element diffusion in our models. 

MESA performs smoothing of certain properties of the stellar interior by default. However, since we want to take into account the chemical gradient in the near-core region and the discontinuities in the Brunt-Väisälä frequency, this behavior is suppressed by setting the following flags:
\begin{itemize}
    \item \texttt{num\_cells\_for\_smooth\_brunt\_B=0},
    \item \texttt{num\_cells\_for\_smooth\_gradL\_composition\_term=0},
    \item \texttt{remove\_mixing\_glitches=False}.
\end{itemize} 

\subsubsection{Mixing}\label{sec:Mixing}
A crucial aspect of stellar modeling involves a sensible prescription for internal mixing, as it plays an important role in a star's lifetime. The convective instability regions are determined using the Ledoux criterion and predictive mixing \citep{Paxton2018}. This method is directly implemented in MESA and allows for a robust determination of the boundaries of the convective zones between each time step, resulting in more reliable model convergence. Proper determination of the convective boundaries is important in the near-core region because it affects the mixing as the core recedes, and hence the Brunt-Väisälä frequency \cite{Gabriel2014}, which is fundamental for the extraction of oscillation frequencies in GYRE. No semi-convection or thermohaline convection is considered. We treat convection using mixing length theory \citep{BohmVitense1958}, specifically the method described by \citet{Cox1968}, which is implemented in MESA. The mixing length parameter is set to the typical value of $\alpha_{\rm{MLT}} = 1.80$.

MESA allows for different implementations of convective core overshooting. The most commonly used are step overshoot and exponential overshoot \citep[see e.g.][]{Pedersen2018}. In the former case, the size of the convective core is essentially extended by a fraction of its size. In the latter case, the mixing efficiency is assumed to decay exponentially from the core boundary. \citet{Pedersen2018} provide an overview of the different prescriptions (see their Fig. 1) and found that $g$-modes can help distinguish between the two on the main sequence, though this capability is lost at later stages of evolution. We focus only on the exponential overshooting scheme and set the value for the overshooting parameter to $f_{\rm{ov}} = 0.0175$, according to typical values that can be expected for intermediate-mass $\gamma$ Doradus stars \citep{Mombarg2021}. Although there is evidence for a mass-dependence of the overshoot parameter \citep{Claret2019}, we find that our estimates of critical magnetic field strengths do not depend strongly on the overshooting (see Section \ref{sec:Toys}). This mass-dependence would have the largest impact on the lower stellar masses of our grid, since the overshooting parameter is observed to plateau for higher masses \citep{Claret2019}. Therefore, we keep it constant throughout the grid for simplicity. The exponential decay is active until the diffusion coefficient associated with overshooting drops below the value $\texttt{overshoot\_D\_min} = 0.01$. 

Finally, we include additional diffusive mixing in the stellar envelope, defined by the coefficient $D_{\rm{ext}} = 10\ \rm{cm^2/s}$. This can be used to mimic the mixing induced by, for example, rotation in a highly simplified way, which prolongs the main sequence lifetime. Moreover, there is strong evidence in the literature for the need for envelope mixing to reproduce observations of the period spacing patterns in $g$-modes \citep{Moravveji2015, Moravveji2016}. 

\begin{figure}
    \centering
    \includegraphics[width=\linewidth]{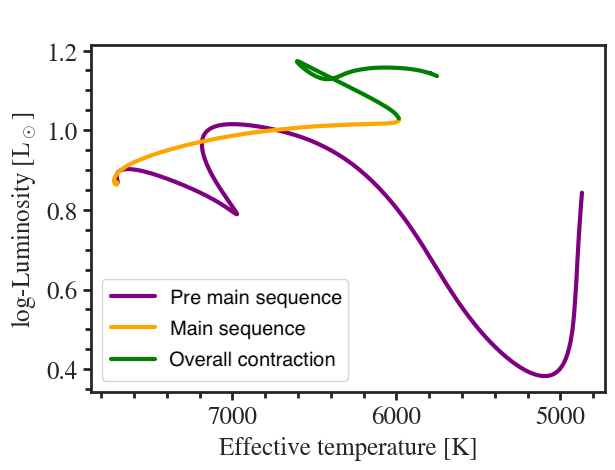}
    \caption{Schematic of the three stages in our MESA models.}
    \label{fig:Evotracks}
\end{figure}

\subsubsection{Resolution and convergence}\label{sec:ResolutionAndConvergence}
Our evolutionary tracks are configured in three stages because the time scales of evolution of the stellar interior are significantly different. We first evolve our stars through the pre-main sequence stage with a maximum time step of 10000 years, terminating the model right before the ignition of hydrogen fusion in the core. The second stage corresponds to the main sequence (MS) lifetime of the star until the hydrogen is depleted in the core. We set a lower limit of 0.001 on the central hydrogen fraction as a stopping criterion. The time resolution is lower with a step size of 1 million years. At the terminal-age main sequence (TAMS), we initialize a third evolutionary stage to evolve the star along the overall contraction (OC) phase, which is much shorter than the main sequence. The time step is, in this case, 100000 years. A schematic of the different stages can be found in Fig. \ref{fig:Evotracks}. The final stage is terminated when the evolutionary tracks cross a lower boundary of $5750 \rm{K}$ in effective temperature. At this point, all models within the considered mass range have completed the 'hook' seen in the HR diagram. For the MS and OC phases, the input physics remain the same, with the only difference being the stopping criteria and temporal resolution. 

The spatial resolution of our models is set by the \texttt{mesh\_delta\_coeff} parameter, which controls the difference between grid points along the stellar profile of the model and typically has a value in the range of 0.2 to 1.2 \citep{Gautam2025}. Low values enforce a fine grid, whereas larger values allow for coarser resolution and faster computation times. Since we model stars with a receding core that leaves behind a sharp chemical gradient and $g$-modes are highly sensitive to this \citep[e.g.][]{Miglio2008}, we require a fine resolution for our model and use $\texttt{mesh\_delta\_coeff} = 0.4$. Smaller values were found to introduce convergence issues in our current configuration. We also enforce a maximum size for the cells with the parameter $\texttt{max\_dq} = 0.001$ \citep{Pedersen2018}, which is given in terms of the total mass of the star.

\subsection{Grid model selection}\label{sec:Grid}
To determine the best model from our grid to model the stars in our sample, we use the Mahalanobis distance, introduced for modeling $g$-modes by \citet{Aerts2018}. We define a set of observables $\theta$ that are used to compare the stars in our sample with the same quantities derived from our model grid. The aim is to find the model with the smallest distance $\theta_0$ to our observations:
\begin{equation}
    \theta_0 = \rm{arg} \rm{min}\ \biggl\langle(\mathbf{Y}_j - \mathbf{Y}^*)^\intercal(V + \Lambda)^{-1}(\mathbf{Y}_j - \mathbf{Y}^*)\biggl\rangle,\ j=1..q,
\end{equation}
where $\mathbf{Y_j}$ corresponds to the observables of the grid point $j$ and $\mathbf{Y^*}$ is the vector containing those of our target star. This calculation is carried out for all models $q$ in our grid. $\rm{V}$ is the covariance matrix of the observables $\theta$ and $\rm{\Lambda}$ is a diagonal matrix containing the squared uncertainties $\sigma_\theta^2$. This method has proven valuable in the analysis of $g$-modes due to the ability to account for correlations between observables \citep[see e.g.][]{Aerts2021, Michielsen2021, Pedersen2021}. We consider three observables, the effective temperature and luminosity of the star as well as the asymptotic period spacing, defined as
\begin{equation}
    \Pi_0 = 2\pi^2\left(\int_{R}\frac{N}{r}dr\right)^{-1}.
\end{equation}
The asymptotic period spacing carries information on stellar age, as it decreases along the evolutionary track of a star. It is therefore useful in constraining the best model from a grid. Our vector thus becomes $\theta = (T_{\rm{eff}}, L_\odot, \Pi_0)$, which is similar to what was done by \citet{Mombarg2019}, where the luminosity was substituted for surface gravity. The choice of observables allows us to determine the best model from our grid simply from the MESA profiles without extracting frequencies with GYRE, thus saving a considerable amount of computation time. In the case of 44 Tau, there is no value for $\Pi_0$. We therefore compute the radial fundamental pressure mode $(\ell,n) =(0,1)$ and its first overtone, and use this to constrain the age of the star, since pressure modes are shifted towards lower frequencies as the star evolves. This gives $\theta = (T_{\rm{eff}}, L_\odot, f_0, f_1)$.

\subsection{GYRE frequency extraction}\label{sec:GYRE}
We extract frequencies with GYRE v.7.2.1 \citep{Townsend2013, Townsend2018}, using as input the computed MESA stellar profiles. A template inlist can be found in \autoref{sec:GYREinlist}. Since we do not consider mode excitation in this work, we focus on adiabatic oscillations. We treat rotation through the Traditional Approximation of Rotation (TAR) \citep{Eckart1960, Berthomieu1978, Townsend2013}, which incorporates the influence of the Coriolis force associated with the horizontal component of the angular velocity on the pulsations. The TAR is valid under the assumption that $\omega \ll N$, where $\omega$ is the wave frequency and $N$ is the Brunt-Väisälä frequency. This applies for low-frequency $g$-modes even in fast-rotating stars, since they are not very sensitive to the centrifugal deformation of the star \citep{Dziembowski2007, Saio2018, Henneco2021}. We use vacuum boundary conditions for pressure perturbation and extract high radial order $g$-modes $n \in [-150; -15]$ for the best model in our grid for the $\gamma$ Doradus stars. Other common choices in boundary conditions do not alter our extracted frequencies significantly In the considered range, $g$-modes are closely spaced in period, therefore requiring a fine frequency grid for GYRE to extract all values. We limit the scanned frequency range to $0.3\rm{-}5 \rm{d^{-1}}$ and find that a grid with 1000 points typically performs well for the stars considered in this work. We then find a correspondence between the modeled and observed oscillations for each $(\ell,m)$ pair by matching the radial orders from GYRE that lie closest to the observed pulsations in frequency. This allows us to interpret our models later on and determine the critical magnetic field strength required to suppress the high-order $g$-modes that are not found in the observations of these stars. 

For 44 Tau, the approach is slightly different, since we are dealing with mixed modes for which modeling is not as straightforward. We scan a range of frequencies defined by the lowest and highest frequency listed in \autoref{tab:44TauSummary} and extract all modes in this range. From here, we determine the best-matching GYRE frequencies in the same way as for the $\gamma$ Doradus stars. We also note that the TAR approximation is valid for the $\ell=2$ mixed modes, since their frequencies are below the Brunt-Väisälä frequency and their radial wavenumber is larger than the horizontal wavenumber. 

\begin{figure}
    \centering
    \includegraphics[width=\linewidth]{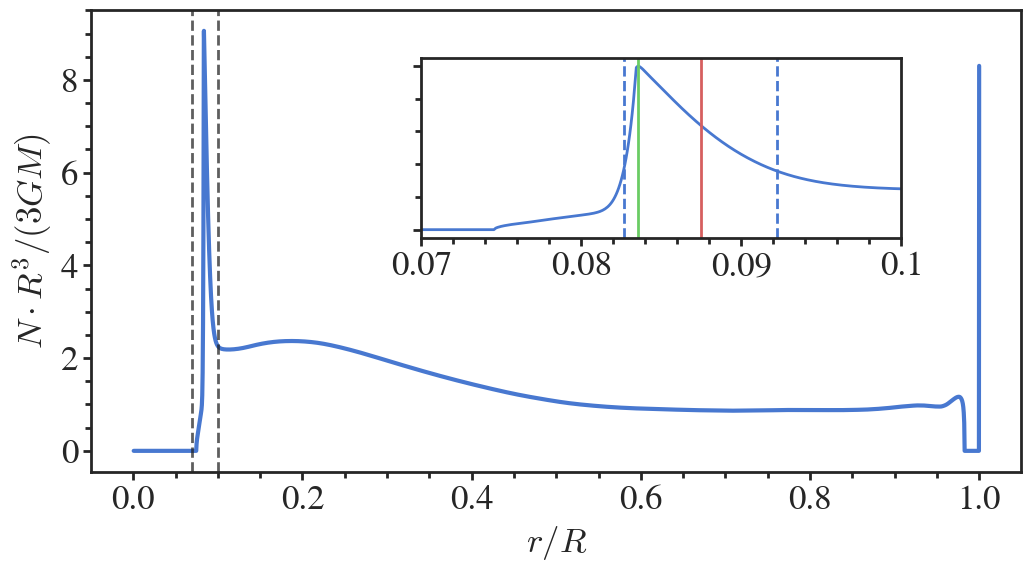}
    \caption{Example Brunt-Väisälä frequency from one of our models as a function of fractional radius, scaled with $R^3/(3GM)$ to be dimensionless. The subplot focuses on the region near the spike of $N$ that is delimited by the gray dashed lines. The inner and outer limits of the spike in the inset plot correspond to the blue dashed lines. The green line highlights the radius of the maximum of the spike, where we solve the eigenvalue problem, while the red line corresponds to the mean radius of the spike.}
    \label{fig:Brunt}
\end{figure}

\subsection{Dedalus eigenvalue problem}\label{sec:Dedalus}
The critical magnetic field strength is then calculated using the Dedalus code \citep{Vasil2019, Lecoanet2019, Burns2020}. As in \citet{Lecoanet2022}, we solve eigenvalue problems to determine if a propagating gravity waves forms a standing wave ($g$-mode) or if its energy is converted to magnetic waves and dissipated. We investigate a range of magnetic field strengths and gravity wave frequencies, allowing us to determine the magnetic field strength required to suppress each $g$-mode extracted from GYRE. Because higher radial orders are suppressed at weaker magnetic fields than lower radial orders, the critical field strength $B_{\rm{crit}}$ associated with the observed mode of the highest order gives an upper limit on the near-core magnetic field strength $B_r$. Since mode suppression occurs in the chemical composition gradient region of the star, where the Brunt-Väisälä frequency spikes in the near-core region, we solve the eigenvalue problem solely in the region of the spike, as shown in Fig. \ref{fig:Brunt}. To isolate the spike, we set a minimal threshold $N\geq10^{-5}\,\rm{Hz}$ for the Brunt-Väisälä frequency to determine the inner and outer limits, shown in dashed blue lines. The green line shows the radius where the spike peaks. We compute the maximum frequency $N$ and stellar density at this radius, which are key quantities for evaluating mode suppression. The red line shows the mean radius of the spike and corresponds to the radius at which we solve the eigenvalue problem. This is done since the exact radius of the spike is not generally well constrained by the MESA models, and fluctuations can affect this estimation. 

The magnetic field configuration assumed in \citet{Lecoanet2022} was that of a dipole field $B_r(r, \theta)=B_r(r)\cos(\theta)$. In our work, we include two additional terms. Firstly, we introduce a quadrupole component to the poloidal field as the $\ell=2$, $m=0$ spherical harmonic:
\begin{equation}
    B_{r, \mathbb{Q}}(r, \theta)=B_{r, \mathbb{Q}}(r)(3\cos^2(\theta) -1)\, ,
\end{equation}
and the total radial field thereby becomes: 
\begin{equation}
    \begin{split}
        B_r(r, \theta) &= B_{r, \mathbb{D}}(r, \theta)+B_{r, \mathbb{Q}}(r, \theta) \\
        & = B_r(r)\cos(\theta) + B_{r, \mathbb{Q}}(r)(3\cos^2(\theta) -1)\, .
        \label{eq:Poloidal}
    \end{split}
\end{equation}
Secondly, we implement a toroidal magnetic field of the form:
\begin{equation}
    B_{\phi, \mathbb{T}}(\theta)=B_{\phi, \mathbb{T}}\sin(\theta)\, . 
    \label{eq:Toroidal}
\end{equation}
This toroidal field is axisymmetric and satisfies the regularity conditions at the poles.

In spherical coordinates, the near-core magnetic field is therefore written as the sum of three components, such that:
\begin{equation}\begin{split}
    \boldsymbol{B}(r, \theta)&= B_r(r,\theta)\,\boldsymbol{e_r}+B_\phi(\theta)\,\boldsymbol{e_\phi} \\
    &=\left(B_{r, \mathbb{D}}(r, \theta)+B_{r, \mathbb{Q}}(r, \theta)\right)\,\boldsymbol{e_r} + B_{\phi, \mathbb{T}}(\theta)\,\boldsymbol{e_\phi} \\ 
    &= \left(C_1B_{r, \mathbb{D}}(r)\cos(\theta)+C_2B_{r, \mathbb{Q}}(r)(3\cos^2(\theta) -1)\right)\,\boldsymbol{e_r} \\  &+ C_3B_{\phi, \mathbb{T}}\sin(\theta)\,\boldsymbol{e_\phi}
    \label{eq:Br}
\end{split}\end{equation}
where $B_{r, \mathbb{D}}(r)$, $B_{r, \mathbb{Q}}(r)$, and $B_{\phi, \mathbb{T}}$ represent the magnetic field strengths of the dipole, quadrupole, and toroidal components of the field, respectively. $C_1$, $C_2$, and $C_3$ are normalization constants that are defined such that:
\begin{equation}
    \int \boldsymbol{B}^2\,d\Omega=1.
    \label{eq:Normalization}
\end{equation}

Solving Eq. \ref{eq:Normalization} for each component leads to the following normalization:
\begin{equation}
    C_1=\sqrt{\frac{3}{4\pi}}\, ; \, C_2=\sqrt{\frac{5}{16\pi}};\,C_3 =\sqrt{\frac{3}{8\pi}}
\end{equation}

The modifications to the implemented magnetic field lead to a change in the linearized horizontal momentum and induction equations, Eqs. 4 and 6 of \citet{Lecoanet2022}. The oscillation equations we solve for are now defined as follows:
\begin{equation}
    ik_ru_r+\frac{1}{r}\nabla_h\boldsymbol{u_h}=0 ,
    \label{eq:Continuity}
\end{equation}
\begin{equation}
    \frac{N^2}{\omega}u_r+k_rp=0,
    \label{eq:Radial}
\end{equation}
\begin{equation}\begin{split}
    -i\omega\boldsymbol{u}_h+2\boldsymbol{\Omega}\times\boldsymbol{u}_h+\frac{1}{r}\boldsymbol{\nabla}_hp&=\frac{i\,B_{r, \mathbb{D}}(r)}{4\pi\rho_0}\cos(\theta)\,k_r\boldsymbol{b}_h \\
    &+\frac{i\,B_{r, \mathbb{Q}}(r)}{4\pi\rho_0}(3\cos^2(\theta)-1)\,k_r\boldsymbol{b}_h \\
    &+\frac{B_{\phi,\mathbb{T}}}{4\pi\rho_0r}\sin(\theta)\boldsymbol{e}_\phi\boldsymbol{\cdot}\boldsymbol{\nabla}_h\boldsymbol{b}_h,
\end{split}\label{eq:MagneticField}\end{equation}
\begin{equation}
    \begin{split}
    -i\omega\boldsymbol{b}_h & -B_{\phi,\mathbb{T}}\sin(\theta)\boldsymbol{e}_\phi\boldsymbol{\cdot}\boldsymbol{\nabla}_h\boldsymbol{u}_h \\
    & -i\left(B_{r, \mathbb{D}}(r)\cos(\theta)+B_{r, \mathbb{Q}}(r)(3\cos^2(\theta)-1)\right)k_r\boldsymbol{u}_h = 0.
    \label{eq:Lecoanet2}
    \end{split}
\end{equation}

Here, $k_r$ is the radial wavenumber that we solve for, $\omega$ is the oscillation frequency, $r$ is the radius of the spike in the Brunt-Väisälä frequency, $p$ is the pressure, $\nabla_h$ denotes the angular gradient, $\boldsymbol{\Omega}$ is the rotation vector, $\rho_0$ is the density, $u_r$ is the radial velocity, and $\boldsymbol{u}_h$ and $\boldsymbol{b}_h$ are the horizontal velocity and magnetic field vectors, respectively. In Eq. \ref{eq:Lecoanet2}, we neglect the Ohmic damping term present in Eq. 6 of \citet{Lecoanet2022} and consider the ideal case where $\eta = 0$. The continuity equation, Eq. \ref{eq:Continuity}, and the radial momentum equation, Eq. \ref{eq:Radial}, remain unchanged. These oscillation equations are valid in the WKBJ approximation under the assumption that the toroidal component of the magnetic field is much stronger than the poloidal magnetic field, $B_h/B_r\sim N/\omega$, which is expected \citep[e.g.,][]{Ratnasingam2024, Takata2026}.

To determine which eigenvalues are numerically converged, we solve for the eigenvalues for two different spatial resolutions. We consider, respectively, 128 and 192 spherical harmonics in the low and high resolution grids. The relative difference between the eigenvalues from both grids is:
\begin{equation}
    \Delta k_r = \frac{k_{r, \rm{lres}} - k_{r, \rm{hres}}}{k_{r,\rm{lres}}},
\end{equation}
where $k_{r,\rm{lres}}$ and $k_{r, \rm{hres}}$ represent the numerical value of $k_r$, computed from the low-resolution and high-resolution grids, respectively. Our condition for convergence is that $\Delta k_r$ is less than $10^{-7}$. We vary the value of the field $B$ until there is no converged eigenvalue with the specified frequency $\omega$, implying that the mode is suppressed. This corresponds to the critical magnetic field strength $B_{\rm{crit}}$. We present in Appendix \ref{sec:DedalusEq} the equations of the eigenvalue problem that are written into Dedalus. 

We note here that the mixed modes in 44 Tau are of significantly lower radial order than the $g$-modes observed in the $\gamma$ Doradus stars. Since the WKBJ approximation is only justified for short wavelengths, care must be taken to ensure our method can be applied to this star. From our MESA models, we find that the pressure scale height, $H_p=0.062\,\rm{cm}$, at the spike in the Brunt-Väisälä frequency is larger than the longest wavelength of the $\ell=2$ modes, $\lambda_{\ell=2}=0.015\,\rm{cm}$. The WKBJ approximation is thus valid, and we can compute the critical field strength using Dedalus. We ignore the $\ell=1$ mixed modes in this work since the condition is only marginally satisfied, and the TAR does not apply for these pressure-dominated modes. 

The modifications made to Eqs. \ref{eq:MagneticField} and \ref{eq:Lecoanet2} allow us to determine the effect of different field configurations on the suppression of gravity modes. We will consider a range of different scenarios. Firstly, following \citet{Lecoanet2022}, we will look into a purely poloidal field, i.e., with $B_{\phi, \mathbb{T}}=0$, and investigate three field geometries: 
\begin{itemize}
    \item A purely dipolar ($\mathbb{D}$) field, where $B_{r, \mathbb{Q}}=0$
    \item A purely quadrupole ($\mathbb{Q}$) field, where $B_{r, \mathbb{D}}=0$
    \item A mixed ($\mathbb{M}$) field, where we enforce that the dipole and quadrupole components contribute equally to the total field strength, $B_{r, \mathbb{D}}/B_{r} = B_{r, \mathbb{Q}}/B_r = 0.5$.
\end{itemize}
Secondly, we will include the influence of the toroidal component and examine how its presence affects the calculations above.

In summary, our methodology involves four stages. We first perform mode identification based on observations, which in this case was guided by prior work from \citet{Li2020}. Secondly, we determine the best stellar model for each target using MESA, from which we compute the radius, stellar density, and the value of the spike in the Brunt-Väisälä frequency near the stellar core. Thirdly, for the given structural model, we extract oscillation frequencies with GYRE and match these to the observed modes in order to determine a set of model-specific frequencies. Finally, the MESA and GYRE parameters are used to determine the critical magnetic field strength using Dedalus, from which we can infer an upper limit on the near-core magnetic field strength by considering the value $B_{\rm{crit}}$ of the highest radial order $g$-mode. 

\section{Results}\label{sec:Results}
We report in this section the results obtained from our analysis of the three targets presented in Section \ref{sec:SectionI}. One of our objectives is to assess the consistency of the method for different oscillation and magnetic field geometries. We first introduce the outcome of our analyses for the three stars, assuming a dipole magnetic field. Secondly, we look into the effect of adopting a dipole, quadrupole, or mixed magnetic field configuration. We then investigate how our results change when introducing a toroidal component to the magnetic field. Lastly, we consider some toy scenarios where the initial composition and overshooting parameters are adjusted to investigate the influence of certain MESA parameters. 

\begin{figure}
    \centering
    \includegraphics[width=\linewidth]{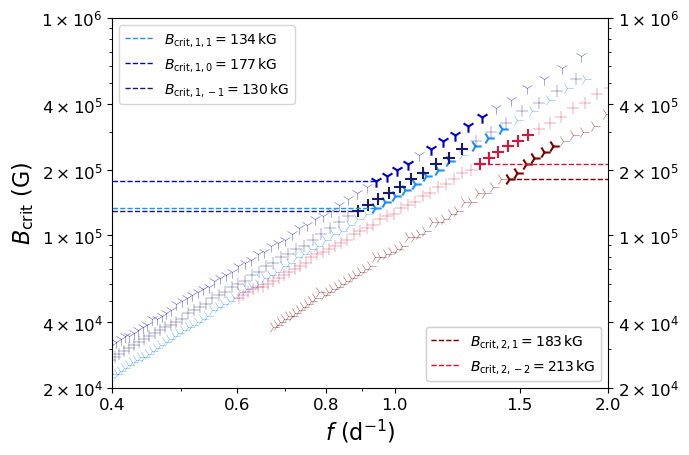}
    \caption{Critical magnetic field strengths, $B_{\rm{crit}}$, for KIC~3127996 as a function of oscillation frequency, assuming a purely dipolar magnetic field. Dipole modes $m=-1, 0, 1$ are shown in blue, while quadrupole modes $m=-2, -1, 1$ are shown in red. Both axes are in logarithmic scale, revealing a linear trend. The order of the $g$-modes increases from the top right corner to the bottom left corner. Bold crosses represent observed frequencies for the star. Thin crosses correspond to higher or lower radial $g$-modes extracted in GYRE. The dashed lines serve to guide the eye on the critical magnetic field strength, $B_{\rm{crit}}$. Each mode has an inferred magnetic field strength, required to suppress the highest observed order. The exact values being provided in the legends.}
    \label{fig:Bcrit_KIC3127996}
\end{figure}

\subsection{Critical magnetic field strengths}\label{sec:BcritResults}
Using the stellar parameters from each target, we find the best matching model from our grid of models. These are listed in \autoref{tab:TargetSummary}. We find mostly consistent values for the stellar parameters within their respective uncertainties, although the effective temperature for KIC~3127996 may be slightly underestimated in our model. Regardless, we consider these best models as good approximations of the stellar interiors, allowing us to determine internal magnetic field strengths. 

Fig. \ref{fig:Bcrit_KIC3127996} shows the magnetic field strength required to suppress a mode as a function of frequency for our best model of KIC~3127996. This serves as an upper limit for the near-core magnetic field strength. We note that all values are within 40\% of each other, with estimates of $B_r$ increasing when using $\ell=2$ modes. We select the lowest value of $B_{\rm{crit}}$ from Fig. \ref{fig:Bcrit_KIC3127996} as an upper limit for the magnetic field strength since the observed $(l,m)=(1,-1)$ modes would have to be suppressed for larger values. Therefore, $B_r \approx 130\pm40\,\rm{kG}$ in KIC~3127996. More information on the uncertainty is provided in Section \ref{sec:Toys}. Estimations of $B_r$ using prograde or retrograde modes yield similar results. 

\begin{figure}
    \centering
    \includegraphics[width=\linewidth]{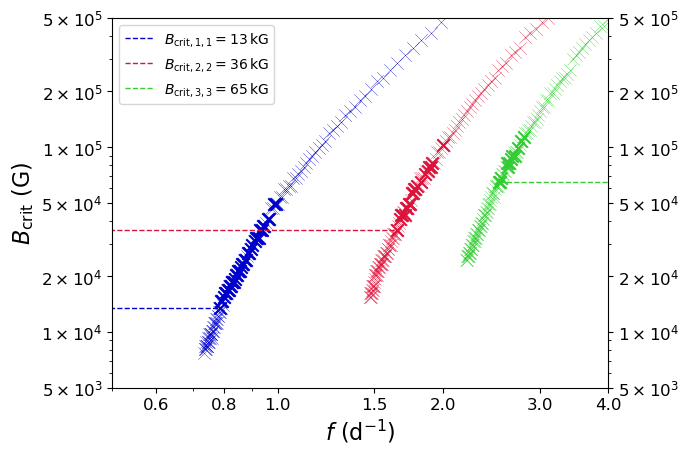}
    \caption{Critical magnetic field strengths, $B_{\rm{crit}}$, for KIC~5876187 as a function of oscillation frequency, assuming a purely dipolar magnetic field. The dipole mode $(l,m)=(1,1)$ is shown in blue, the quadrupole mode $(l,m) = (2,2)$ is shown in red, and the octupole $(l,m)=(3,3)$ mode in green. The data points and dashed lines contain the same information as in Fig. \ref{fig:Bcrit_KIC3127996}.}
    \label{fig:Bcrit_KIC5876187}
\end{figure}

Fig. \ref{fig:Bcrit_KIC5876187} shows the $B_{\rm{crit}}$-diagram for KIC~5876187. The principle is the same as in Fig. \ref{fig:Bcrit_KIC3127996}. We note that, contrary to what is seen in Fig. \ref{fig:Bcrit_KIC3127996}, the relation between $B_{r}$ and the observed frequency is no longer linear in log-space due to KIC~5876187 rotating faster. There is a larger deviation in the values for $B_{\rm{crit}}$ from higher degree modes, with the estimate from $\ell=3$ differing from that using $\ell=1$ by a factor of almost 5. We find that $B_r \approx 13\pm 4\,\rm{kG}$ for KIC~5876187.

\begin{figure}
    \centering
    \includegraphics[width=\linewidth]{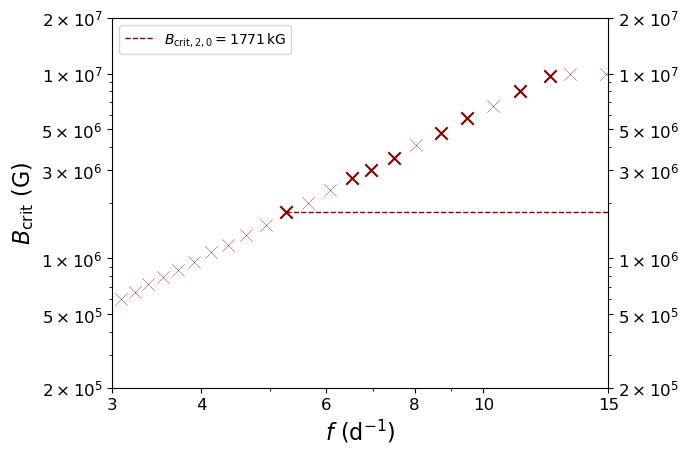}
    \caption{Critical magnetic field strengths, $B_{\rm{crit}}$, for 44 Tau as a function of oscillation frequency, assuming a purely dipolar magnetic field. The quadrupole mode $(l,m) = (2,0)$ is shown in red. The data points and dashed lines contain the same information as in Fig. \ref{fig:Bcrit_KIC3127996}.}
    \label{fig:Bcrit_44tau}
\end{figure}

Fig. \ref{fig:Bcrit_44tau} shows the critical field strength for 44 Tau. Fewer modes are observed for this star, and are of lower order than those in the $\gamma$ Doradus stars. $B_{\rm{crit}}$ is considerably larger in this case than for the other two targets, since we assume in this scenario that all higher-order $g$-modes or mixed modes are suppressed under the influence of the magnetic field. We conclude that $B_r\approx1.8\pm0.5\,\rm{MG}$.

\subsection{Magnetic field configurations}\label{sec:MagCon}
\begin{figure}
    \centering
    \includegraphics[width=\linewidth]{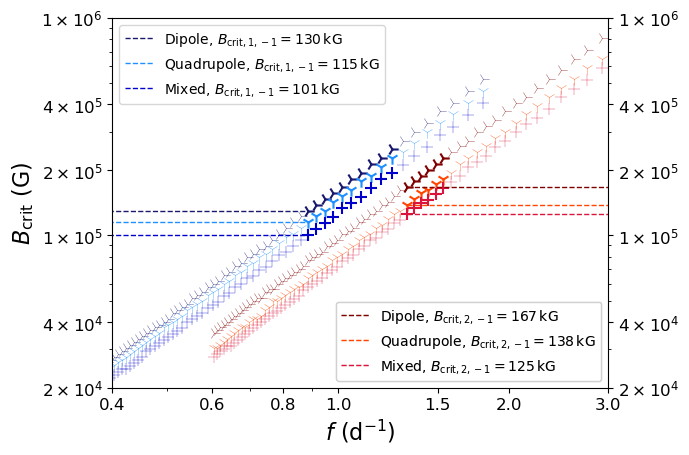}
    \caption{Critical magnetic field strength as a function of frequency for KIC~3127996, for the three different field structures presented in Section \ref{sec:Dedalus}. Results using $(\ell,m)=(1,-1)$ are shown in blue, whilst those using $(\ell,m)=(2,-1)$ are in red. The field structures are color-coded in different shades of the same color for each mode degree.}
    \label{fig:MagCon_KIC3127996}
\end{figure}

\begin{figure}
    \centering
    \includegraphics[width=\linewidth]{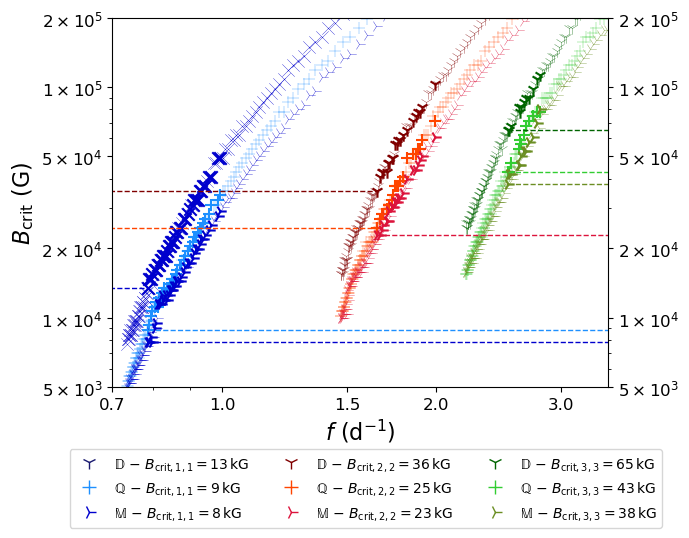}
    \caption{Critical magnetic field strength as a function of frequency for KIC~5876187, for the three different field structures presented in Section \ref{sec:Dedalus}; dipole $(\mathbb{D})$, quadrupole $(\mathbb{Q})$, and mixed $(\mathbb{M})$. Results using $(\ell,m)=(1,1)$ are shown in blue, $(\ell,m)=(2,2)$ in red, and $(\ell,m)=(3,3)$ in green.}
    \label{fig:MagCon_KIC5876187}
\end{figure}

\begin{figure}
    \centering
    \includegraphics[width=\linewidth]{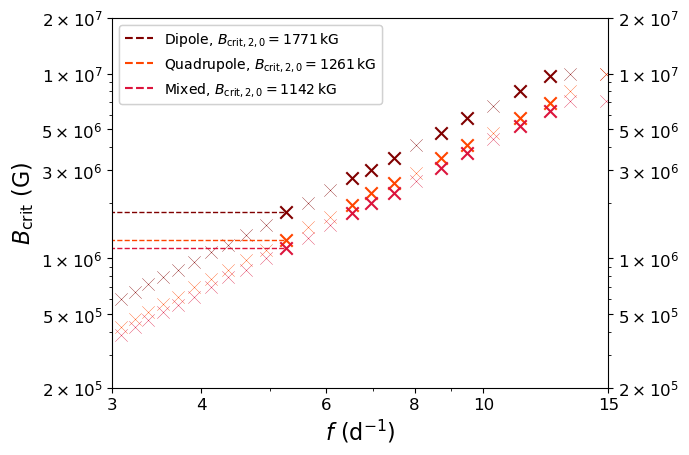}
    \caption{Critical magnetic field strength as a function of frequency for the $\ell=2$ modes in 44 Tau, for the three different poloidal field structures presented in Section \ref{sec:Dedalus}.}
    \label{fig:MagCon_44Tau}
\end{figure}

\begin{figure*}
    \centering
    \begin{minipage}{0.33\textwidth}
        \centering
        \includegraphics[width=\linewidth]{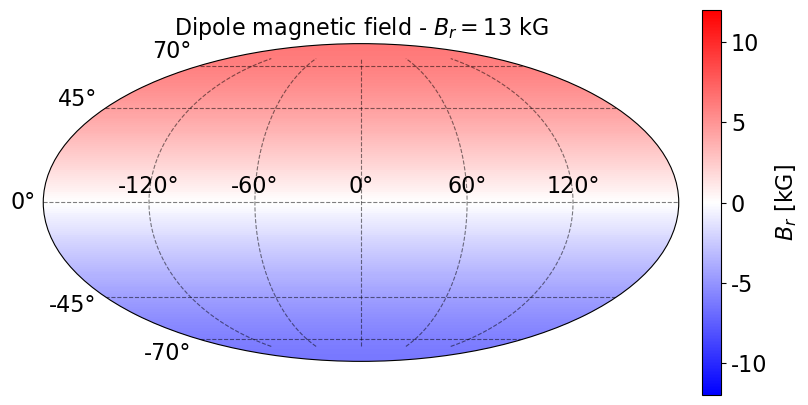}
    \end{minipage}
    \hfill
    \begin{minipage}{0.33\textwidth}
        \centering
        \includegraphics[width=\linewidth]{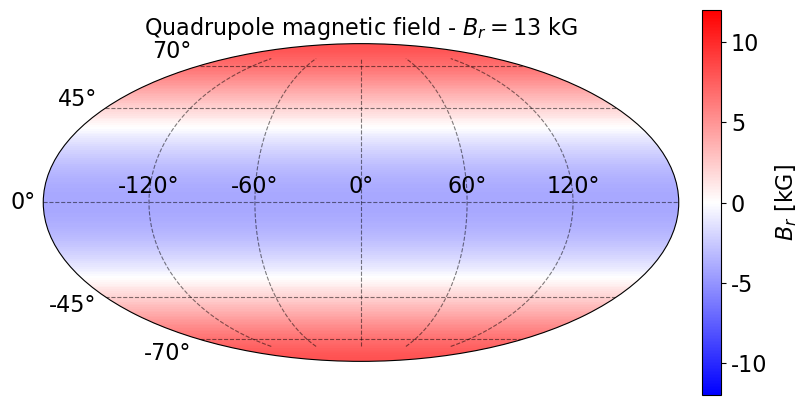}
    \end{minipage}
    \begin{minipage}{0.33\textwidth}
        \centering
        \includegraphics[width=\linewidth]{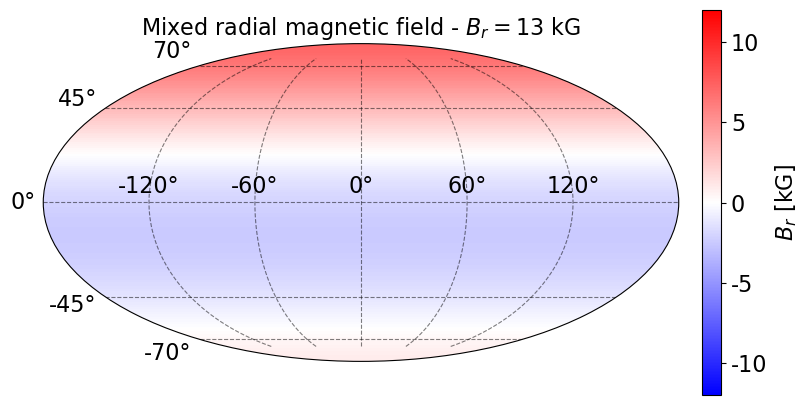}
    \end{minipage}
    \\
    \begin{minipage}{0.33\textwidth}
        \centering
        \includegraphics[width=\linewidth]{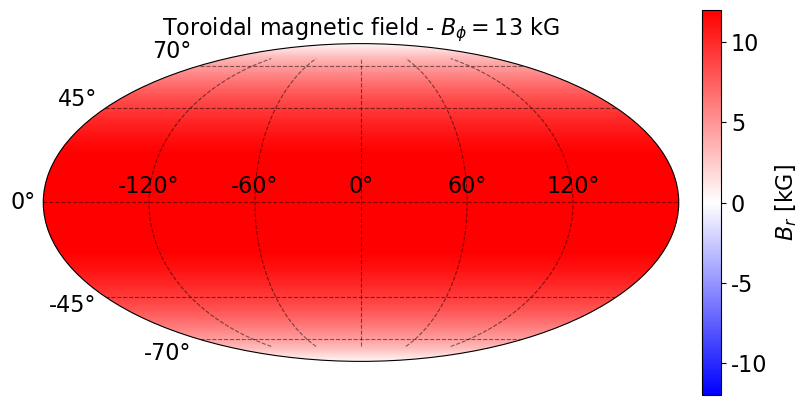}
    \end{minipage}
    \begin{minipage}{0.33\textwidth}
        \centering
        \includegraphics[width=\linewidth]{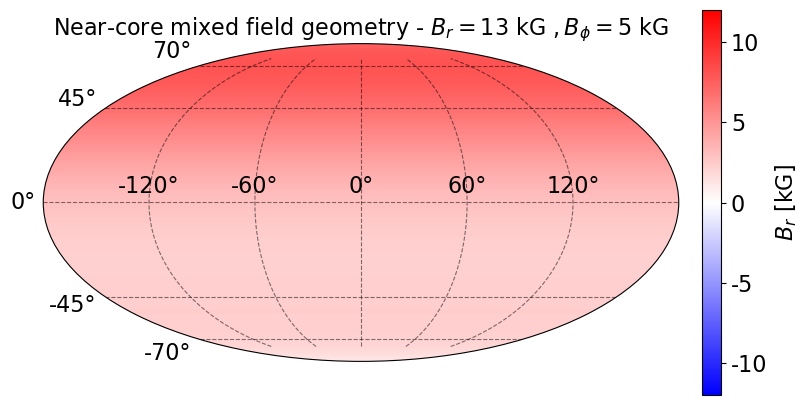}
    \end{minipage}
    \caption{Radial component  $B_r$ of the near-core magnetic field for KIC~5876187, computed from Eq. \ref{eq:Br} as a function of latitude $\theta$ and longitude $\phi$. \textit{Upper row:} from left to right, the panels show a dipole, quadrupole, and mixed field configuration. The symmetry of the field in the latter case is broken due to the constructive and destructive interference of the dipole and quadrupole components. The strength of the magnetic fields are equal to the critical fields inferred from Fig. \textit{Lower row:} from left to right, we show the purely toroidal field and mixed field with contributions from all three terms $B_{r,\mathbb{D}}, B_{r,\mathbb{Q}}$ and $B_{\phi,\mathbb{T}}$.}
    \label{fig:MagViz}
\end{figure*}

We present in Figs. \ref{fig:MagCon_KIC3127996}-\ref{fig:MagCon_44Tau} the estimates of the critical magnetic field strengths using the three different poloidal field configurations mentioned in Section \ref{sec:Dedalus} for KIC~3127996, KIC~5876187, and 44 Tau. We consider for each star the mode geometry for which we find the upper limit for $B_r$ in the previous section. We note that weaker magnetic fields are required to achieve similar results when considering quadrupole or mixed fields. For KIC~5876187, we provide the projections of $B_r$ on a sphere in Fig. \ref{fig:MagViz}, for the three different field configurations. 

For our three targets, we compute the fractional difference in $B_{r}$ between the quadrupole field and the dipole and mixed fields, respectively:
\begin{equation}
     \delta B_{r\, \mathbb{Q}} = 1-\frac{B_{r,\, \mathbb{Q}}}{B_{r,\,\mathbb{D}/\mathbb{M}}},
\end{equation}
which helps us determine if a mode is more sensitive to specific field configurations. The values are shown in Table \ref{tab:MagFrac}. We note that the $\ell=2$ modes do not appear to be more sensitive to quadrupole or mixed field configurations than the $\ell=1$ modes. This characteristic is consistent for all stars. No robust conclusion can be drawn for $\ell=3$ modes, which are only observed in KIC~5876187. For the three stars, we note that the critical field strengths associated with the quadrupole field and the mixed field are closer to each other compared to that of the dipole field. All estimates remain within 70\% and 90\% of that of the dipole field. 

\begin{table}[]
    \centering
    \caption{Fractional differences in $B_r$.}
    \begin{tabular}{ccc|ccc|ccc}
         \hline
         \multicolumn{3}{c|}{KIC 3127996} & \multicolumn{3}{c|}{KIC 5876187} & \multicolumn{3}{c}{44 Tau} \\
         $\ell$ & $\mathbb{Q}/\mathbb{D}$ & $\mathbb{Q}/\mathbb{M}$ & $\ell$ & $\mathbb{Q}/\mathbb{D}$ & $\mathbb{Q}/\mathbb{M}$ & $\ell$ & $\mathbb{Q}/\mathbb{D}$  & $\mathbb{Q}/\mathbb{M}$ \\
         \hline
         $1$ & 0.85 & 0.86 & $1$ & 0.69 & 0.86 &  &  &  \\
         $2$ & 0.83 & 0.90 & $2$ & 0.69 & 0.91 & $2$ & 0.71 & 0.91 \\
         & & & $3$ & 0.72 & 0.88 & & & \\
         \hline
    \end{tabular}
    \tablefoot{We show values depending on the assumed field configuration, for all mode degrees observed in our stars. The first column for each target shows the degree $\ell$. The second and third columns contain the fractional difference of the dipole critical field strength relative to that of the quadrupole and mixed fields, respectively.}
    \label{tab:MagFrac}
\end{table}

If we assume that the magnetic field can be of dipole nature and that it scales with distance such that $B_r \propto r^{-3}$, we can extrapolate the strength of the field at the surface of the star. All relevant quantities can be found in Table \ref{tab:BcritValues}. For KIC~3127996, a dipole field of $B_r \approx 130\,\rm{kG}$ would have a strength of $B_{\rm{surf}} \approx 76\,\rm{G}$ at the surface. With a quadrupole field of strength $B_r\approx115\,\rm{kG}$, scaling as $B_r \propto r^{-4}$, the surface field would be $B_{\rm{surf}} \approx 6\,\rm{G}$. Similar estimates for KIC~5876187 and 44 Tau are found in Table~\ref{tab:BcritValues}. 
While the near-core magnetic field required in our modeling of 44 Tau is considerably stronger than the $\gamma$ Doradus stars, the surface field remains weak. 

\subsection{Inclusion of a toroidal magnetic field}\label{sec:ToroidalFieldInclusion}
As previously mentioned, toroidal magnetic fields are expected to be stronger than the poloidal components of the near-core magnetic fields \citep{Takata2026}. We therefore include a formulation for the toroidal field in the oscillation equations and modify our computations to assess its role in relation to gravity mode suppression. For each star, we freeze the poloidal component to some fixed value below the estimated $B_{\rm{crit}}$ and solve the eigenvalue problem for the highest observed gravity mode for a range of toroidal field strengths to determine at what $B_{\mathbb{T}, \rm{crit}}$ the mode is suppressed. In the case of KIC~3127996, for a dipole field $B_r=100$~kG, the suppression of the mode only occurs for a toroidal field $B_{\phi,\mathbb{T}}\approx20\,\rm{MG}$. We recall that the critical field strength in the purely poloidal case was $B_r\approx130\,\rm{kG}$. The strength of the toroidal field is 200 times stronger than the radial component.  This behavior is consistent for prograde and retrograde modes, different spherical degrees, and different poloidal field configurations, as we find the same order of magnitude for $B_{\phi, \mathbb{T}}$ in all cases. An even larger ratio is found for KIC~5876187, where $B_{\phi, \mathbb{T}}\approx6\,\rm{MG}$ is required to suppress the highest-order $g$-mode with a poloidal field of $B_r=10\,\rm{kG}$ compared to the critical value of $13\,\rm{kG}$.

\subsection{Model dependencies}\label{sec:Toys}
To evaluate the impact of model parameters on our results, we consider two modified models of KIC~5876187. Firstly, we look at how the initial composition affects the outcome of our analysis by decreasing the initial helium fraction while keeping the metallicity constant. This allows us to examine how a change in the hydrogen content affects the composition gradient near the core, where we determine the critical field strength. The hydrogen fraction is set as $X = 1 - Y - Z$, where $Z = 0.02$ is the metal fraction and $Y = 0.28$ is the default value for the helium fraction. We decrease this value to 0.26 and thereby set the hydrogen fraction to $X = 0.72$. Secondly, we change the value for the overshooting to assess the dependence of our results on the choice of overshooting. We set $f_{\rm{ov}} = 0.001$, effectively reducing overshooting by an order of magnitude, from the default value of 0.0175. In both cases, we proceed similarly to what was done previously, by determining the best model for the stellar parameters of KIC~5876187 along the evolutionary track with a modified initial composition or overshooting parameter.

We show the obtained $B_{r}$ for $(\ell, m)=(1,1)$ for each scenario in Table \ref{tab:BcritValues}. These are obtained by investigating the $B_{\rm{crit}}$ diagrams for these cases, exactly as shown in Figs. \ref{fig:Bcrit_KIC3127996}$-$\ref{fig:Bcrit_44tau}. We retrieve similar values for $B_r$ in both cases, with fractional differences typically around 15\%. When reducing overshooting, the shape of the spike in the Brunt-Väisälä frequency at the core boundary has a larger amplitude. This leads to a smaller $B_r\approx11\,\rm{kG}$. For a large reduction in overshooting, the impact on $B_r$ remains small, suggesting that the method does not depend on tight constraints of the overshooting parameter. When varying the initial composition, we are able to retrieve a model that lies within the same region of the HR diagram for which the Brunt-Väisälä frequency does not vary significantly. Therefore, we obtain closer estimates for the critical field strength, as seen in Table \ref{tab:BcritValues}.

Since our range of observed orders for the $\gamma$ Doradus stars differs from those reported by \citet{Li2020}, we also examined the influence of mode identification on the uncertainties in our $B_{\rm{crit}}$ values by deriving the critical magnetic field strength based on their values. As shown in Table \ref{tab:BcritValues} for $(\ell, m)=(1,-1)$, our estimates using slightly lower-order modes are larger than those obtained from the data provided in \citet{Li2020}, which is consistent with the fact that stronger fields are required to suppress decreasing orders of $g$-modes \citep{Lecoanet2022}. For KIC~3127996 and KIC~5876187, we generally observe a difference in the highest-order mode of $\Delta n_{\rm{pg}}\approx5$ with \citet{Li2020}, resulting in uncertainties in the critical magnetic field strength of roughly $\Delta B_{\rm{crit}}\approx 25\%$. The relative error drops to roughly 10\% if the mode identification differs only by 2 radial orders. We treat the uncertainties from mode identification and modeling parameters as independent, and therefore apply an uncertainty $\sigma_{B_r}\approx30\%$ to our $B_r$ estimates in Section \ref{sec:BcritResults}.

\begin{table}[]
    \centering
    \caption{Summary of magnetic field strengths for all three targets in our sample.}
    \begin{tabular}{c|ccc}
         \hline
         KIC~5876187 & $B_{r, \mathbb{D}}$ & $B_{r, \mathbb{Q}}$ & $B_{r, \mathbb{M}}$ \\
         $B_r$ & $13\,\rm{kG}$ & $9\,\rm{kG}$ & $8\,\rm{kG}$ \\
         $X_{\rm{init}} = 0.72$ & $14\,\rm{kG}$ & $9\,\rm{kG}$ & $8\,\rm{kG}$ \\
         $f_{\rm{ov}} = 0.001\, \rm{cm}^2s^{-1}$ & $11\,\rm{kG}$ & $8\,\rm{kG}$ & $7\,\rm{kG}$ \\
         $n_{\rm{min}}\,;n_{\rm{max}}$ & $12\,\rm{kG}$ & $8\,\rm{kG}$ & $7\,\rm{kG}$ \\
         $B_{r, \rm{surf}}$ & $6\,\rm{G}$ & $0.3\,\rm{G}$ & - \\
         \hline
         KIC~3127996 & $B_{r, \mathbb{D}}$ & $B_{r, \mathbb{Q}}$ & $B_{r, \mathbb{M}}$ \\
         $B_r$ & $130\,\rm{kG}$ & $115\,\rm{kG}$ & $101\,\rm{kG}$ \\
         $f_{\rm{rot}} = 0.518 \, \rm{d}^{-1}$ & $138\,\rm{kG}$ & $105\,\rm{kG}$ & $89\,\rm{kG}$ \\
         $n_{\rm{min}}\,;n_{\rm{max}}$ & $95\,\rm{kG}$ & $87\,\rm{kG}$ & $76\,\rm{kG}$ \\
         $B_{r, \rm{surf}}$ & $87\,\rm{G}$ & $7\,\rm{G}$ & - \\
         \hline
         44 Tau & $B_{r, \mathbb{D}}$ & $B_{r, \mathbb{Q}}$ & $B_{r, \mathbb{M}}$ \\
         $B_r$ & $1771\,\rm{kG}$ & $1261\,\rm{kG}$ & $1142\,\rm{kG}$ \\
         $B_{r, \rm{surf}}$ & $182\,\rm{G}$ & $7\,\rm{G}$ & - \\
         \hline
    \end{tabular}
    \tablefoot{The second, third and fourth columns show the estimated critical magnetic field assuming a dipole, quadrupole, or mixed field structure. The rows $B_r$ and $B_{r, \rm{surf}}$ contain to the near-core field strengths and the corresponding extrapolated surface fields, respectively. The toy cases described in Section \ref{sec:Toys} are listed in the remaining rows. The parameter that is varied compared to the default scenario shown in $B_r$ is described in the first column.}
    \label{tab:BcritValues}
\end{table}

\subsection{Rotational modulation}\label{sec:Modulation}
\begin{figure}
    \centering
    \begin{minipage}{0.24\textwidth}
        \centering
        \includegraphics[width=\linewidth]{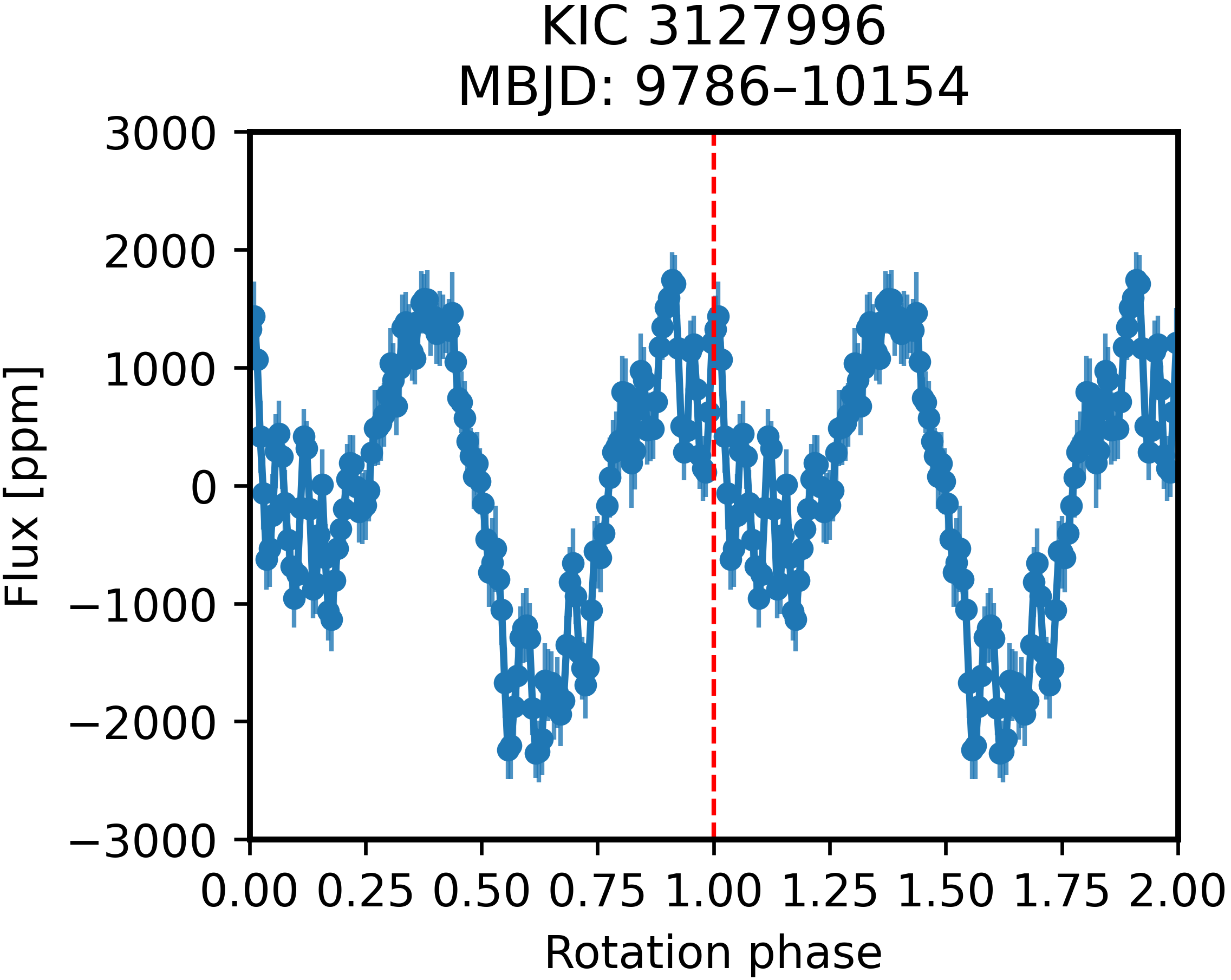}
    \end{minipage}
    \hfill
    \begin{minipage}{0.24\textwidth}
        \centering
        \includegraphics[width=\linewidth]{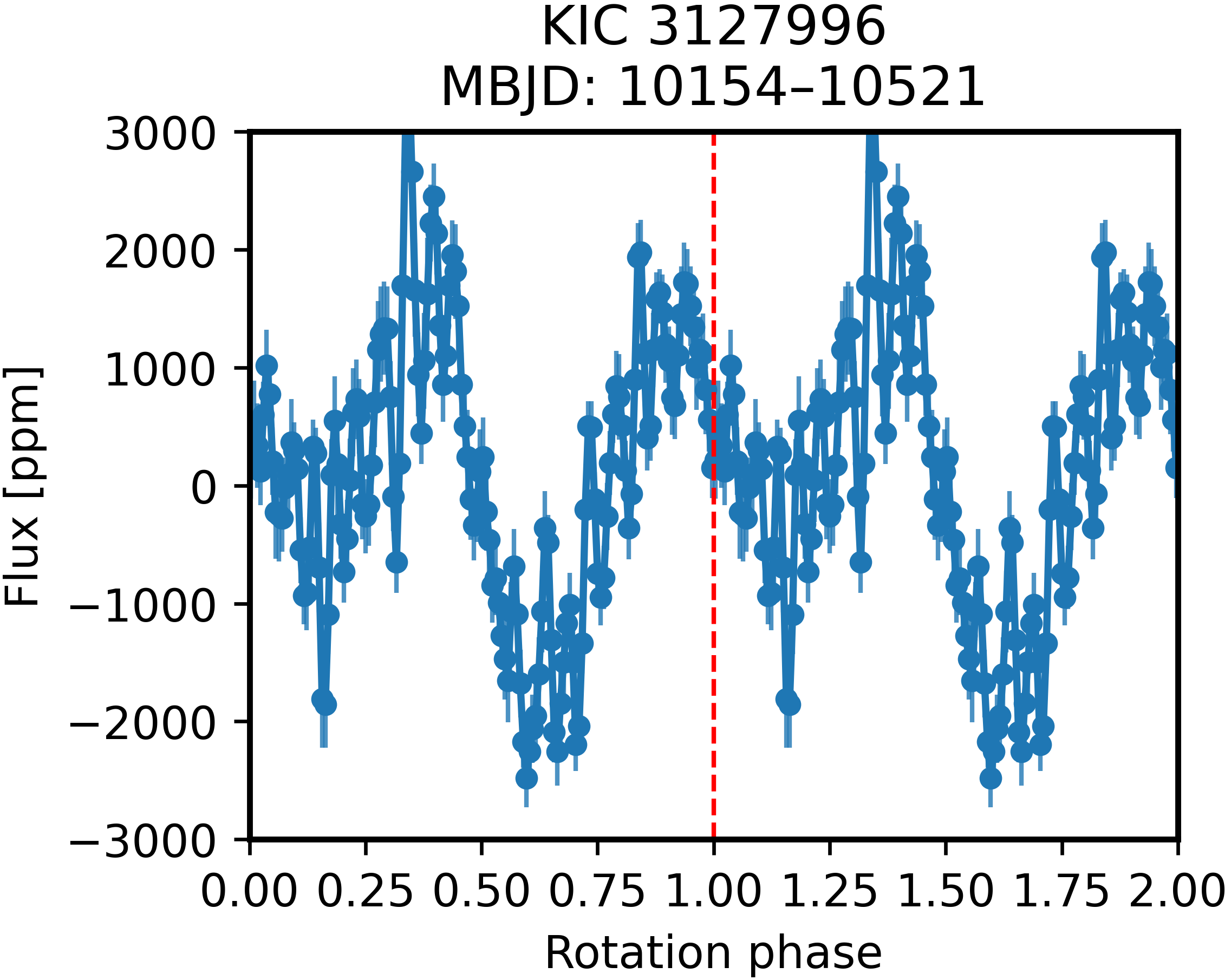}
    \end{minipage}
    \vspace{0.5em}
    \begin{minipage}{0.24\textwidth}
        \centering
        \includegraphics[width=\linewidth]{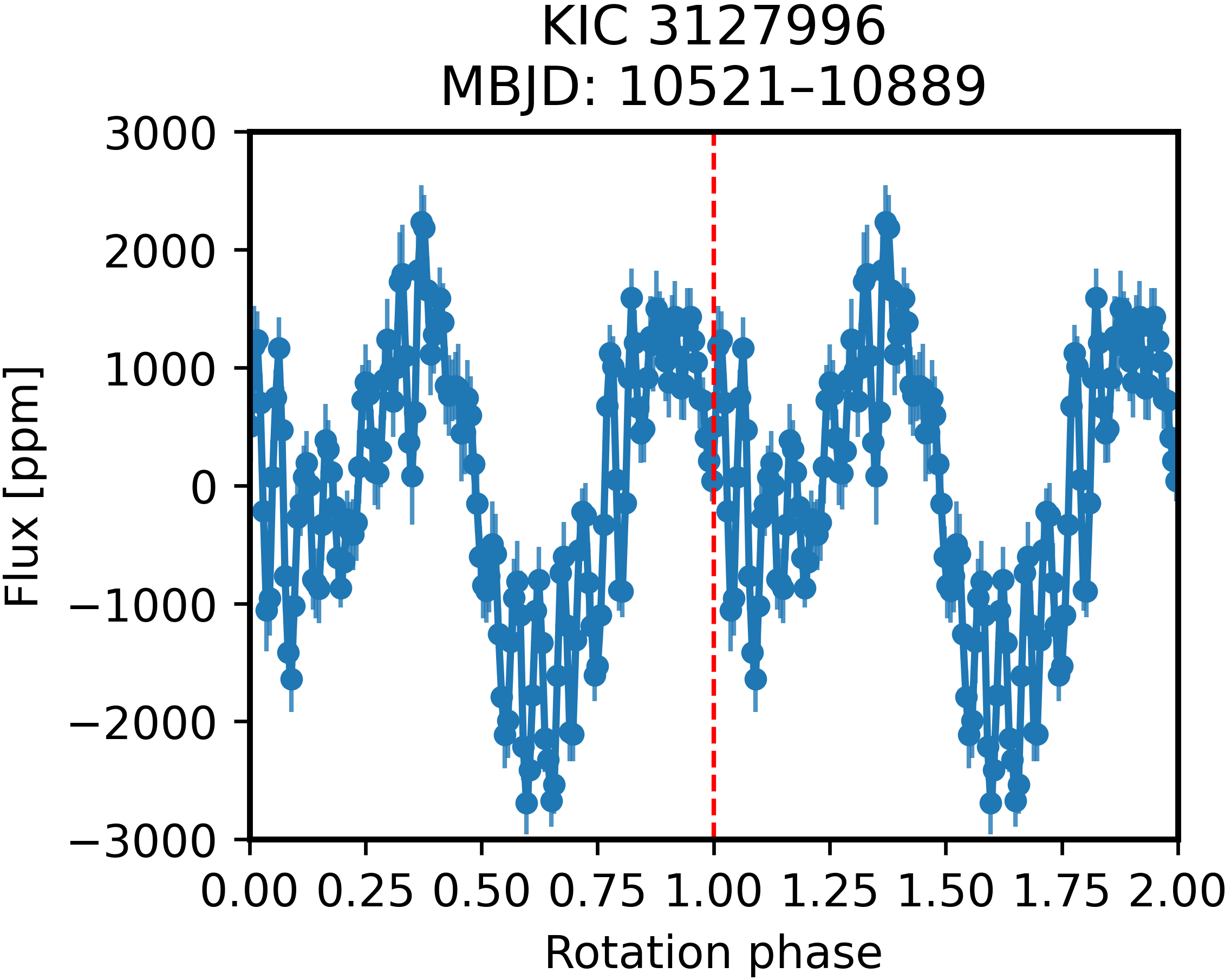}
    \end{minipage}
    \hfill
    \begin{minipage}{0.24\textwidth}
        \centering
        \includegraphics[width=\linewidth]{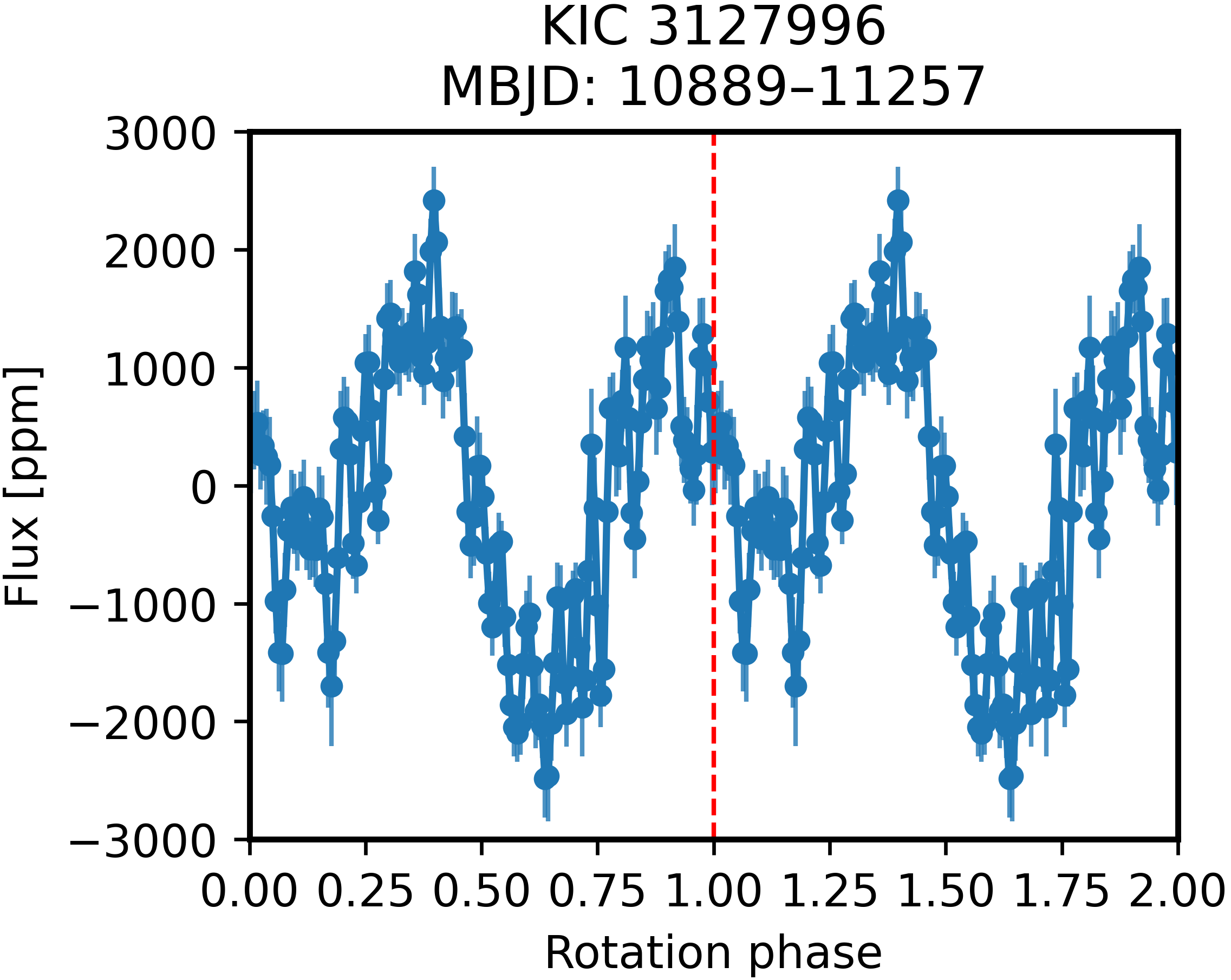} 
    \end{minipage}
    \caption{Phase-folded light curves of KIC~3127996 using \textit{Kepler} data and using the supposed rotation frequency $f=0.0534\,\rm{d^{-1}}$. We subdivided the time series into four separate sections and plot them in separate panels. The MBJD limits for each time series is shown for above the corresponding panel. The features are consistent with rotational modulation.}
    \label{fig:KIC3127996_rotmod}
\end{figure}

We searched for signs of rotational modulation in the light curves of the $\gamma$ Doradus stars in our sample, which would be consistent with the existence of surface magnetic fields. For KIC~3127996, we identified a peak in the amplitude spectrum with frequency $f_{\rm{obs}}=0.0535\,\rm{d^{-1}}$, which is close to the near-core rotation frequency derived from the analysis of $g$-modes (see Table \ref{tab:TargetSummary}). We split the full \textit{Kepler} time series into four segments and phase-folded the light curves with $f_{\rm{obs}}$, as seen in Fig. \ref{fig:KIC3127996_rotmod} to examine possible temporal variations. The data were binned into 150 phase intervals to enhance the signal-to-noise ratio. No clear temporal variability is observed from year to year. 

The shape of the light curve, described by a larger amplitude of the first harmonic compared to $f_{\rm{rot}}$, are consistent with what is seen for other stars showing rotational modulation due to stellar spots \citep[e.g.,][]{Lampens2013, Labadie2023, Antoci2025}. Such variability can also be associated with interactions in a close binary system. However, the binary period of $P=1/f_{\rm{obs}}\approx19\,\rm{d}$ would be too large for such a scenario because tidal distortions would be too weak to cause ellipsoidal variability in a main sequence star. To support this statement, we determine the expected amplitude variation due to ellipsoidal variability as done by \citet{Faigler2011}. We use their Eqs. 2 and 4 and consider a combination of parameters that maximizes the theoretical amplitude for our observed $f_{\rm{rot}}$ and stellar parameters. We assumed a companion star of $1.5\,M_\odot$, such that it is comparable in mass to our target. For KIC~3127996, this results in $A_{\rm{ellip}} \approx 250\,\rm{ppm}$, which is significantly lower than the amplitude variability seen in Fig \ref{fig:KIC3127996_rotmod}. We investigated the Villanova Eclipsing Binary Catalog \citep{Kirk2016} to look for similar signatures in long period binaries, but find no evidence of such ellipsoidal variability at $P_{\rm{orb}}>15\,\rm{d}$. We found five low-resolution spectra from the Large Sky Area Multi-Object Fiber Spectroscopic
Telescope (LAMOST) survey \citep{Zhao2012} with good phase coverage of a potential orbital period, from which we find no evidence for radial velocity shifts, further excluding the possibility of KIC~3127996 being a binary exhibiting ellipsoidal variability. Along with the strong agreement between the observed frequency and the near-core rotation frequency from \citet{Li2020}, there is strong evidence for rotational modulation in KIC~3127996. The small-scale structure in the light curve (Fig.~\ref{fig:KIC3127996_rotmod}) can be associated with a beating effect from one of the observed pulsation frequencies. Band-passing the data while preserving the original sampling, so as to retain only frequencies higher than the rotational frequency and its harmonic, removes the signal from the light curve, confirming that the additional variability originates from these higher frequencies rather than from sampling or other instrumental effects.

\begin{figure}
    \centering
    \begin{minipage}{0.24\textwidth}
        \centering
        \includegraphics[width=\linewidth]{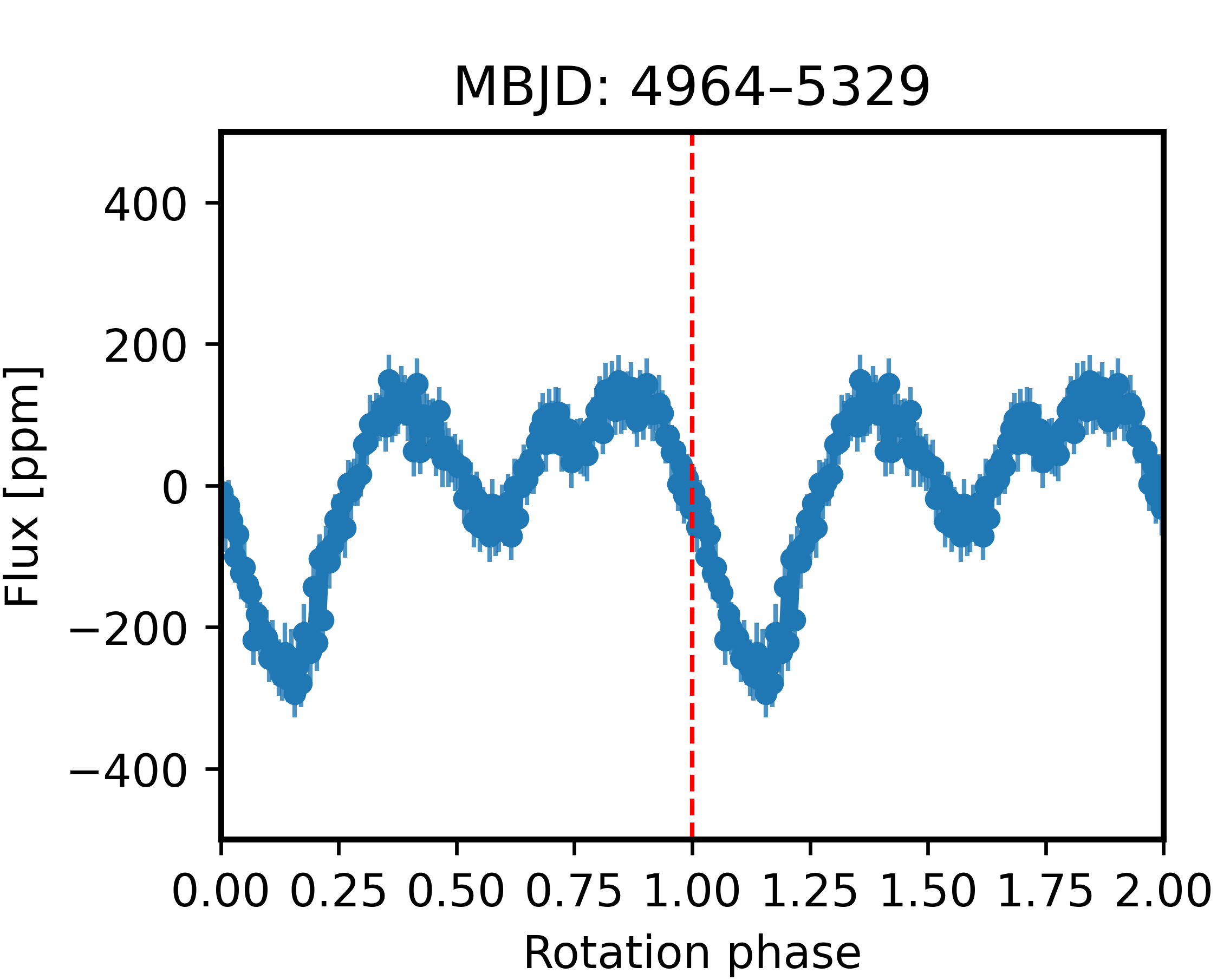}
    \end{minipage}
    \hfill
    \begin{minipage}{0.24\textwidth}
        \centering
        \includegraphics[width=\linewidth]{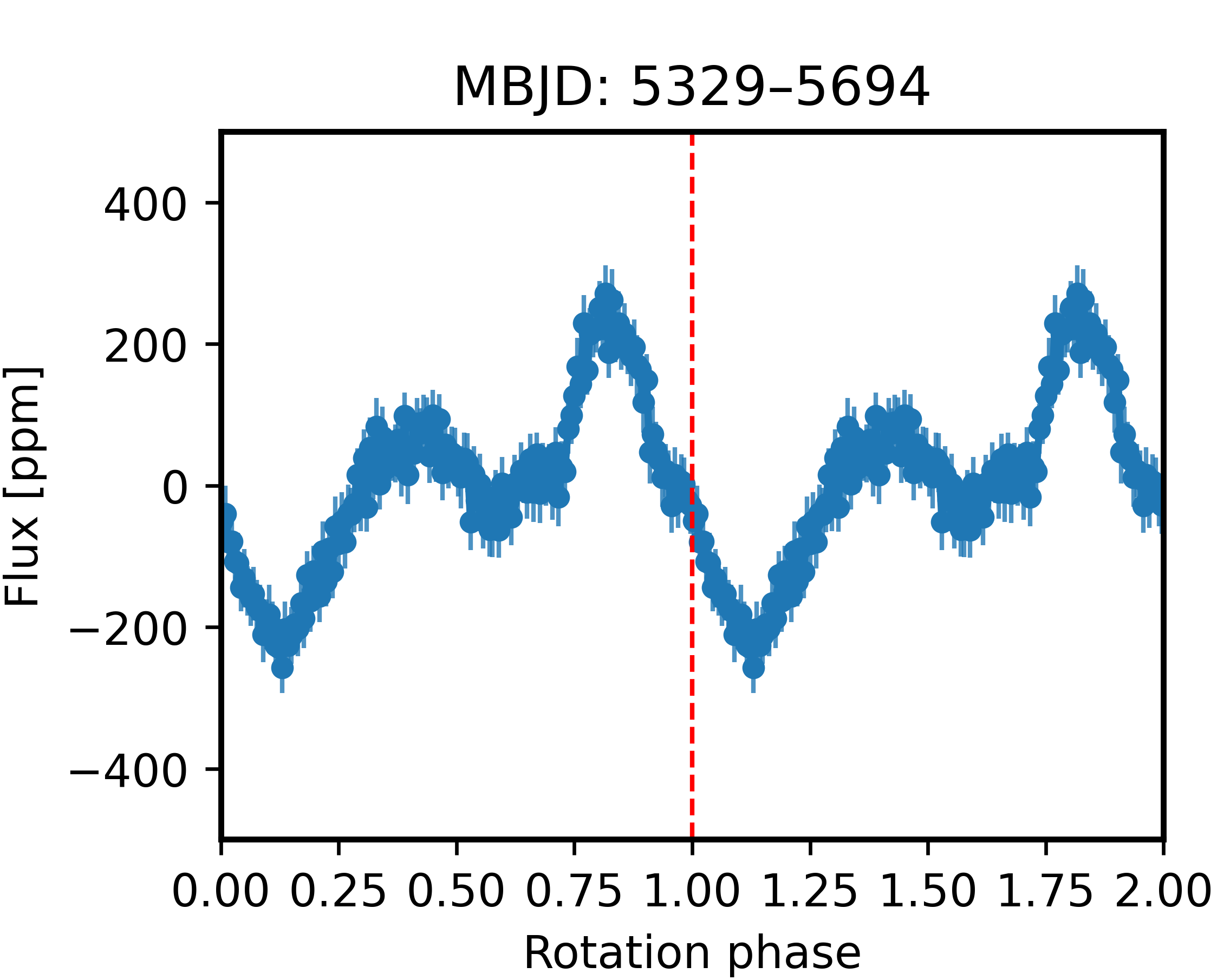}
    \end{minipage}
    \vspace{0.5em}
    \begin{minipage}{0.24\textwidth}
        \centering
        \includegraphics[width=\linewidth]{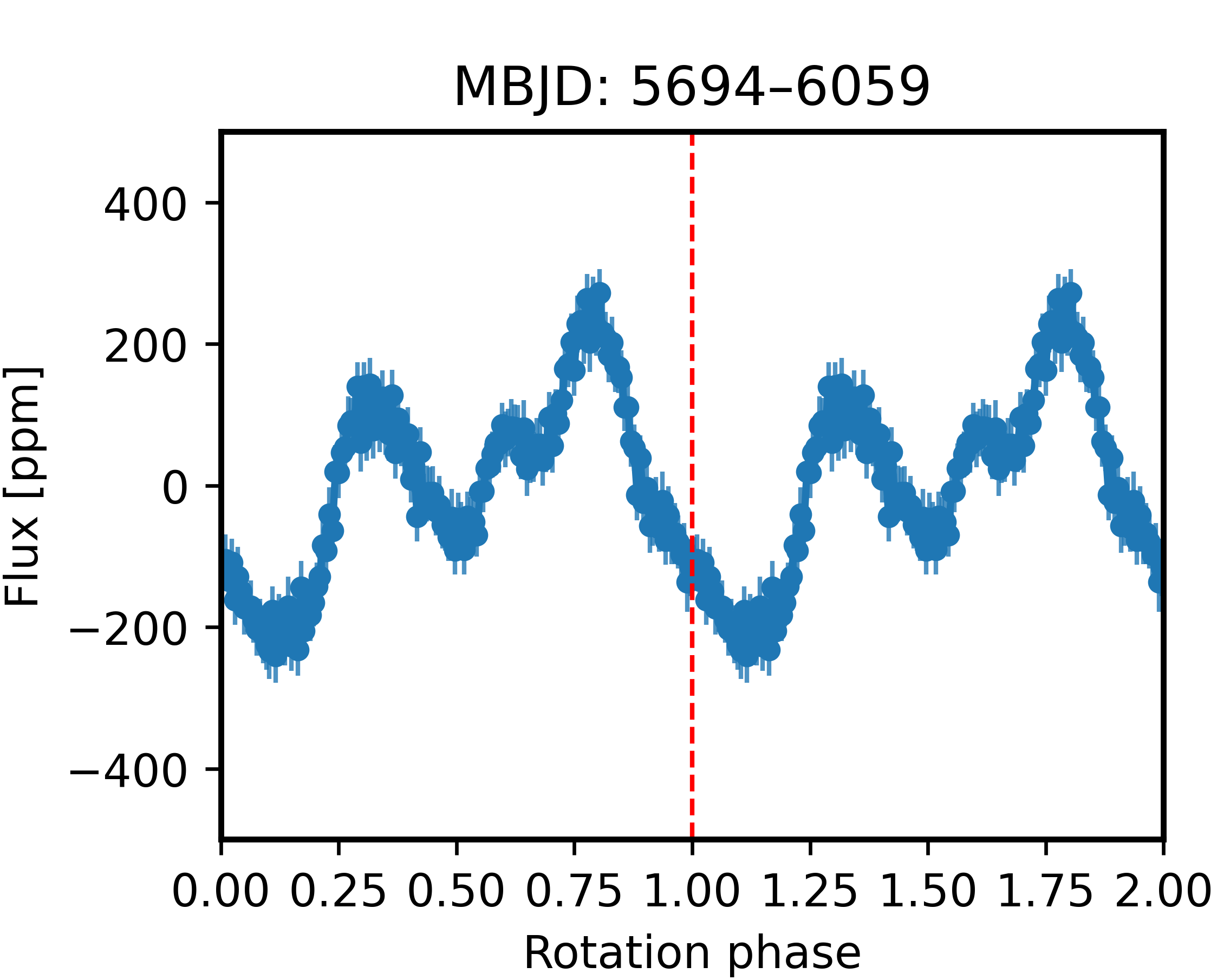}
    \end{minipage}
    \hfill
    \begin{minipage}{0.24\textwidth}
        \centering
        \includegraphics[width=\linewidth]{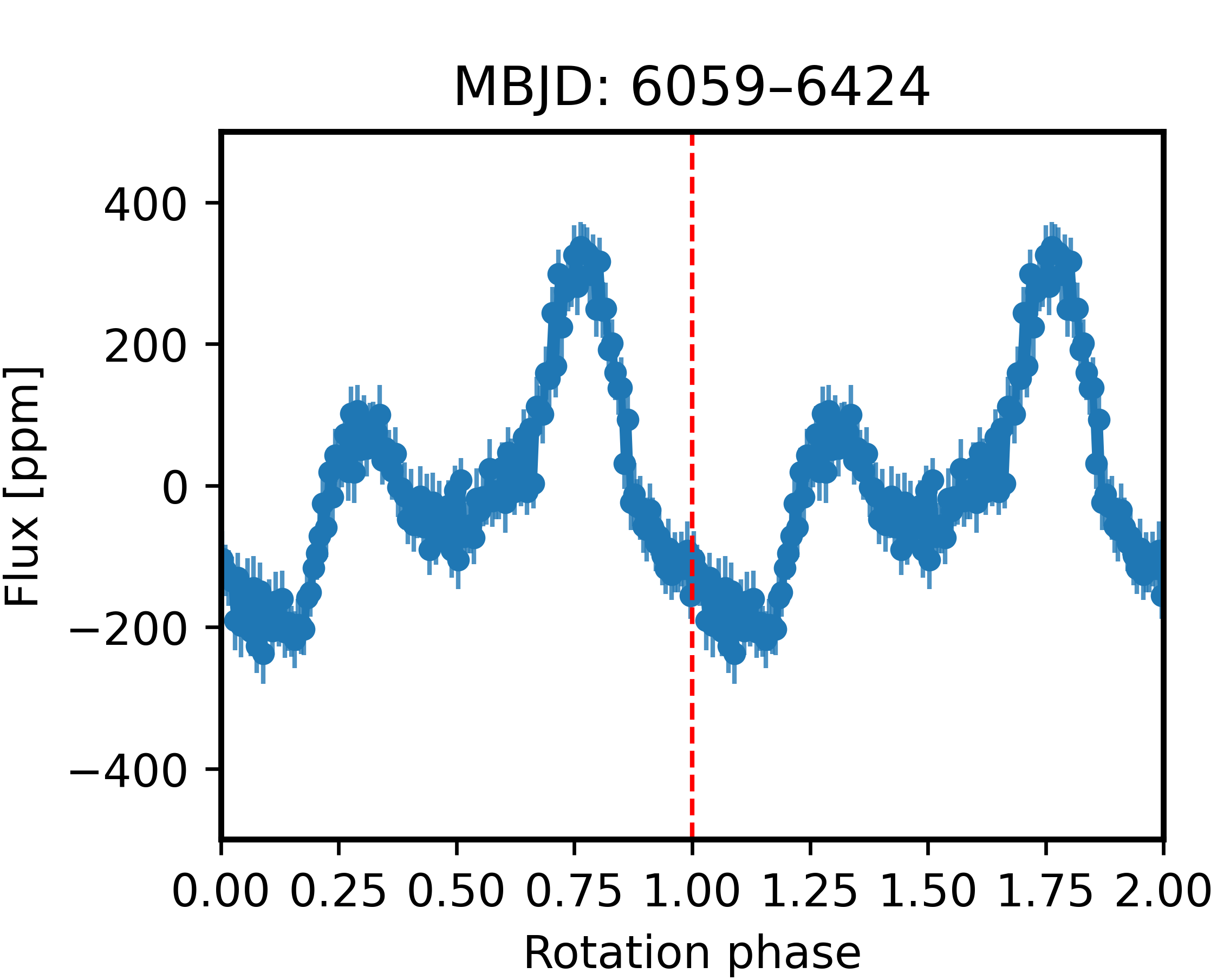} 
    \end{minipage}
    \caption{Same as Fig. \ref{fig:KIC3127996_rotmod}, using the \textit{Kepler} time series for KIC~5876187, with the folding frequency $f=0.583\,\rm{d^{-1}}$. The features are consistent with rotational modulation.}
    \label{fig:KIC5876187_rotmod}
\end{figure}

Similarly, we can match a peak in the amplitude spectrum of KIC~5876187 to the near-core rotation frequency from Table \ref{tab:TargetSummary}. The phase-folded light curve originally contained contributions from the second and third harmonics of the supposed rotation frequency, resulting in a three-maxima structure. However, the third harmonic corresponds to an $(\ell,m)=(2,2)$ $g$-mode that was included in our period spacing pattern to perform mode identification. For this reason, we removed this frequency from the data and computed the phase-folded light curve shown in Fig. \ref{fig:KIC5876187_rotmod}, with $f_{\rm{obs}}=0.583\,\rm{d^{-1}}$. We see the typical structure for rotational modulation and note the large variability between the four different segments, once again supporting the fact that KIC~5876187 likely exhibits rotational modulation. Similarly to what was done for KIC~3127996, the expected amplitude due to ellipsoidal variability is $A_{\rm{ellip}}\approx31000\,\rm{ppm}$, which is several orders of magnitude larger than what we observe in Fig. \ref{fig:KIC5876187_rotmod}. 

As stated above, we observed that a harmonic of the supposed rotation frequency coincides with the frequency of a peak that aligns well with the period spacing pattern for the $\ell=2$ modes. This raises an interesting question about the nature of this peak in the amplitude spectrum. Since the pulsation frequencies we observe are in the inertial reference frame, they are shifted due to the effect of rotation, which suggests a coincidental match. However, we cannot confirm that the peak at the harmonic frequency is an actual $g$-mode, simple harmonic, or a superposition of both.    

In order to estimate the amount of differential rotation, we computed the ratio $f_{\rm{rot, c}}/f_{\rm{rot, obs}}$, where $f_{\rm{rot, c}}$ is the near-core rotation frequency shown in Table \ref{tab:TargetSummary} and $f_{\rm{rot, obs}}$ is the surface rotation frequency derived in Section \ref{sec:Modulation}. All parameters are gathered in Table \ref{tab:Modulation}. From the computed ratio, it can be seen that the near-core region in KIC~3127996 rotates slower than the surface. Although uncommon, this has already been observed for one other $\gamma$ Doradus, KIC~5985441 \citep{Saio2021}. The amount of differential rotation is only on the order of a few percent (see the last column in Table \ref{tab:Modulation}) which is consistent with other measurements of the near-core rotation frequency in $\gamma$ Doradus stars \citep{Li2020}. Similar values are found when analyzing core rotation rates with gravito-inertial modes \citep{Saio2021}. From the observational data, both targets in our sample therefore seem to experience near-uniform rotation, supporting our assumptions when modeling these stars (see Section \ref{sec:SectionI}). 

We note that the two $\gamma$ Doradus stars present in this work were not reported as rotational variables by \citet{Li2020}, whereas we find clear signs of rotational modulation in their light curves. In \citet{Li2020}, the authors found roughly 10\% of their sample to exhibit rotational modulation. The identification of two additional targets with such characteristics could suggest that a larger fraction of $\gamma$ Doradus stars show detectable rotation than what was reported in \citet{Li2020}.

\begin{table}[]
    \centering
    \caption{Rotation frequencies for the $\gamma$ Doradus stars in our sample.}
    \begin{tabular}{c|ccc|c}
         \hline
         Target & $f_{\rm{rot,c}}\, \rm{[d^{-1}]}$ & $f_{\rm{rot, obs}}\, \rm{[d^{-1}]}$ & $f_{\rm{rot,c}}/f_{\rm{rot, obs}}$ & $d_{\rm{rot}}$ \\
         \hline
         KIC~3127996 & 0.0518 & 0.0535 & 0.97 & 3\% \\
         KIC~5876187 & 0.596 & 0.583 & 1.02 & 2\% \\
         \hline
    \end{tabular}
    \tablefoot{The first columns shows the near-core rotation derived from $g$-modes \citep{Li2020}. The second columns contains the estimated surface rotation frequency from rotational modulation. The ratio of the two is listed in the third column and the percentage of differential rotation is shown in the final column.}
    \label{tab:Modulation}
\end{table}

\section{Discussion}\label{sec:Discussion}
This work presents a modified version of the method from \citet{Lecoanet2022} to estimate the critical magnetic field strength, $B_{\mathrm{crit}}$, from observations of intermediate-mass stars, for various magnetic field configurations. In this approach, we assume that the damping of the higher radial-order modes is caused by the magnetic fields. Damping of $g$-modes is, however, also predicted due to radiative damping \citep{Dupret2005}. Although a more refined MESA model could be obtained by varying additional parameters in an attempt to improve the estimates of $B_{\mathrm{crit}}$, this is left for future work. In this section, we discuss the implications of our findings, the potential limitations of our estimates, and additional considerations that we will address in future work. 

\subsection{Impact of mode frequencies}\label{sec:ImpactOfModeFreqs}
Fundamental to our analysis is the correct mode identification of the observed pulsations, especially on the highest-order $g$-mode. We showed in Section \ref{sec:Toys} that a difference of 5 orders on $n_{\rm{max}}$ has an impact on $B_{r}$ of the order of 25\%. Different assumptions on the significance of peaks in the amplitude spectra of a star from one observer to another may therefore lead to small changes in $B_{r}$, although this uncertainty remains acceptable within our framework. 

Additionally, since we do not carry out in-depth modeling of our targets, the resulting GYRE frequencies extracted from our best MESA model may deviate slightly from the observed values. This is most significant for 44 Tau, where we observe differences of the order $0.1\,\rm{d}^{-1}$ between observed and simulated values. This, nonetheless, remains acceptable for a crude modeling of an evolved star. For the $\gamma$ Doradus stars, which are still on the main sequence, the frequencies match significantly better with differences of $\Delta_f\approx 0.005\,\rm{d}^{-1}$ only. These deviations will logically produce small changes in $B_{r}$, since the critical field strength is inherently dependent on the frequency in this method. The impact is, however, small in comparison to the choice of mode order $n_{\rm{pg}}$ or degree $\ell$. 

\subsection{Modeling uncertainties}\label{sec:ModelingUncertainties}
The uncertainties regarding the oscillation frequencies discussed above are only part of the total uncertainty budget in our modeling. These additional aspects will be considered in greater detail in a future paper, although some general comments can be made. 

Since the observed $g$-mode spectrum depends on the structure of $N$, the oscillations we extract from observations can be considered a direct measurement of the Brunt-Väisälä frequency, which is used for our calculations of $B_r$. Our modeling aims to determine a model with an internal profile from which we can extract similar frequencies, placing constraints on the values for $N$, $\rho_0$, and $r$ at the spike of the Brunt-Väisälä frequency in Eqs. \ref{eq:MagneticField} and \ref{eq:Lecoanet2}. In this context, it is important to note that we do not perform any smoothing of the Brunt-Väisälä frequency when computing our stellar models. The smoothing options in MESA influence both the amplitude and width of the spike, reducing the former and increasing the latter (see the red and green lines in Fig. \ref{fig:Brunt}). The width of the spike influences its calculated radius, and therefore also $\rho_0$.

While investigating how changes to the assumptions in MESA affect the determination of the critical field strengths, we showed in Section \ref{sec:Toys} and Table \ref{tab:BcritValues} that the strength of the field is of the same order in all toy models presented. This suggests that the method remains robust to small changes in these parameters. The Brunt-Väisälä frequency varies strongly with evolution, implying that constraints on parameters that serve as proxies for stellar age, such as $\Pi_0$ or $f_0$ (the radial fundamental mode, see Section \ref{sec:Grid}), are more important for retrieving an appropriate structure of $N$. However, the effect of overshooting and chemical composition should be explored in more detail, since these parameters can affect the spike of the Brunt-Väisälä frequency, and consequently also the outcome of the computations in Dedalus. Variations in metallicity may result in evolutionary tracks for stars of significantly different stellar masses overlapping in the HR diagram \citep[e.g.,][]{Pamyatnykh2000}. This, in turn, implies that effective temperature and luminosity cannot distinguish between two models if no other parameters are used during model selection. Additionally, work by \citet{Mombarg2019} suggests that the mixing length parameter $\alpha_{\rm{MLT}}$ has an influence on $\Pi_0$ and the stellar parameters, which are in this work the three input observables used to determine our best models. Studying the influence of $\alpha_{\rm{MLT}}$ will be useful in the future. It is therefore clear that considering more free parameters will be an important step in future work to allow for thorough grid-based modeling. Additionally, the smoothness of the Brunt-Väisälä frequency profile acts as the buoyancy glitch and can be probed by using the Fourier transform of g-mode period spacings \citep{Guo2026}. This may be useful to constrain internal mixing and age in future modeling attempts.

\subsection{Interpretation of results}\label{sec:InterpretationOfResults}
Fundamental to this work is the assumption that all modes below the critical frequency associated with the highest order $g$-mode are converted to magnetic waves and are subsequently damped. \citet{Dupret2005} showed that radiative damping increases with radial order and is expected to produce a cutoff of $g$-modes at high radial orders. As a result, the critical magnetic field strengths inferred from the presence of observed modes should be interpreted as upper limits on the near-core magnetic field. In the case of KIC~5876187, the absence of high-radial-order $\ell = 3$ $g$-modes cannot be explained solely by magnetic suppression, since the near-core magnetic field strength of approximately 13~kG inferred from the $\ell = 1$ modes would be insufficient to suppress these oscillations. A possible way to distinguish between radiative damping and magnetic suppression is to examine whether a consistent upper limit on the radial magnetic field strength is obtained for different mode families, since magnetic suppression would affect all modes in a similar manner, whereas radiative damping depends sensitively on radial order. Although magnetic fields also modify radiative energy transport and may indirectly influence radiative damping, a detailed treatment of this coupling is beyond the scope of this work. Nevertheless, our estimates of $B_r$ provide upper limits on the near-core magnetic field strength, as they represent the maximum field compatible with the observed $g$-modes. Ultimately, modeling the excitation and damping of the modes is required to assess whether radiative damping alone can account for the absence of the observed $\ell = 3$ modes.

We provide estimates of the magnetic field strengths at the surface in Table \ref{tab:BcritValues}, assuming that the field has relaxed into a current-free configuration. In this case, a dipolar field scales with radius as $B_r\approx r^{-3}$, while a quadrupolar field scales as $B_r\approx r^{-4}$. Since there has been no detection of magnetic fields in $\gamma$ Doradus stars through observations for a detection threshold of the order of 20-100 G (\citet{Bagnulo2006}; Labadie-Bartz et al. in review), these values remain consistent with weak surface fields in such targets. Furthermore, our estimates of $B_r$ for the $(\ell, m) =(1,1), (1,-1)$ pairs are in agreement with the fact that the effect of the magnetic field on pulsations in the co-rotating frame is the same for all modes of a multiplet \citep{Deheuvels2024}, such that $B_{r, m=1}\approx B_{r, m=-1}$. Spectropolarimetry could possibly allow us to determine the surface magnetic field of 44 Tau, since it is a bright star ($V\approx 5.4$). However, no observations have been carried out to date. KIC~3127996 and KIC~5876187 are quite faint, with Gaia magnitudes $G>11$, and it is therefore challenging to detect and confirm the existence of surface magnetic fields for these targets. Under our assumptions, KIC~3127996 and KIC~5876187 appear to host internal magnetic fields of significantly different strengths. KIC~3127996 is slightly less evolved and slowly rotating, with a magnetic field roughly ten times stronger compared to KIC~5876187. This is an interesting contrast in light of the core dynamo field scenario, where faster rotation should increase the field strength \citep{Brun2005, Augustson2016}. However, we assume that all magnetic fields are aligned with the rotation and pulsation axes. The field obliquity may affect results significantly and must be included in future work.

In more evolved stars, the radial component of near-core magnetic fields has been estimated for red giants from asymmetries in mixed mode splittings and the deviations caused in period spacing patterns, with values typically ranging from 25 to 150 kG \citep{Li2022, Li2023} but also extending up to 600 kG \citep{Deheuvels2023}. Supposing that these fields were formed during the main sequence of the stars from a core dynamo, magnetic flux conservation allowed the authors to predict the field strength during the main sequence. They find initial strengths of dipole, axisymmetric fields in the range of 1 to 3 kG \citep{Li2022}, extending up to 40 kG for the strongest cases \citep{Deheuvels2023}. Based on these results, our estimate for the magnetic field strength in KIC~5876187 could be consistent with the magnetic fields observed in red giant stars. 

On the other hand, a significantly larger core field has also been inferred for the red giant KIC~8561221, showing that highly magnetized red giants can have radial field strengths $B_r>1\,\rm{MG}$ \citep{Fuller2015}. If these were to originate from a core dynamo on the main sequence, this would require $B_r\gg100\,\rm{kG}$. Our analysis of $\ell=2$ mixed modes in 44 Tau could be consistent with such a scenario. However, as previously mentioned, magnetic field suppression is not the only explanation for the absence of higher order mixed modes or $g$-modes. Further detailed modeling of this target would be necessary to determine if such an internal magnetic field is realistic. 3D MHD simulations indicate that intermediary field strengths of several hundreds of kG can be expected in main sequence intermediate-mass stars, with core dynamo fields possibly ranging from 30 to 60 kG \citep{Brun2005, Hidalgo2024}. Additionally, simulations by \citet{Featherstone2009} and more recently \citet{Hidalgo2025} have shown that core dynamos can be strengthened by a fossil field. In the former case, the fields can reach up to 500 kG, while typical values are found in the range of 100-150 kG in the latter case. The stronger field in KIC~3127996 might correspond to a strengthened core dynamo through interaction with a fossil field. This was also suggested by \citet{Lecoanet2022} to explain their results for HD 43317, where a surface field of $B=1312\pm332\,\rm{G}$ \citep{Briquet2013, Buysschaert2017} was detected and shown to be consistent with a fossil scenario.

We introduced a toroidal component in our magnetic field calculations and find that for a poloidal field strength slightly under the critical value $B_{\rm{crit}}$ required for g-mode suppression, the toroidal magnetic field needs to be 200 times stronger to suppress the same frequency. The ratio is significantly larger than $B_\phi/B_r\approx26$ as derived for the magnetic field detected in another $\gamma$ Doradus star \citep{Takata2026}. 3D MHD simulations also suggest that the toroidal component can be stronger than the radial component in the near-core region \citep{Featherstone2009, Hidalgo2024, Ratnasingam2024}, although there is a wide variety of allowed ratios $B_\phi/B_r\approx2\rm{-}100$. Analytical work carried out by \citet{Barrault2025} suggests that toroidal fields can reach up to several MG in the near-core region. It appears that our required toroidal field strength to explain the observed range of $g$-modes in KIC~3127996 and KIC~5876187 is higher than what is expected. Since $g$-modes are mostly transverse in nature, the suppression is governed by the radial magnetic field and our results are therefore insensitive to toroidal fields. As stated previously, the derived values for $B_r$ and $B_\phi$ are only upper limits and suggest that a wide range of toroidal field strengths, consistent with estimates from the previous works mentioned above, can exist simultaneously with a reasonable poloidal field capable of suppressing the high-order $g$-modes. To better understand these aspects, future work will entail accurately modeling KIC~9244992, the star studied by \citet{Takata2026}, and computing critical magnetic field strengths with Dedalus. This will allow us to determine if we can reproduce the detected magnetic field with the observed $g$-mode frequencies and offers a unique opportunity to compare our methods.

In this work, we model stars as rigid rotating bodies. MHD simulations suggest that core dynamo magnetic fields smooth out differential rotation within the convective cores of intermediate-mass stars \citep{Brun2005, Featherstone2009}. From observations of $g$-modes, uniform rotation has also been inferred in the envelope of stars \citep{Li2020}. In some $\gamma$ Doradus stars where $g$-modes and inertial modes couple, core and envelope rotation rates were measured and allowed to determine a near-uniform rotation throughout the stars \citep{Saio2021}. This is also found in KIC~9244992, where \citet{Takata2026} provide the detection of a strong toroidal magnetic field along with a radial component. Approximating our targets as solid-body rotators therefore seems reasonable, although we cannot actually confirm this is the case for 44 Tau in the absence of high-order $g$-modes. 3D MHD simulations show that a substantial shear can occur at the spike of the Brunt-Väisälä frequency near the convective core, contributing to stable mixed poloidal-toroidal magnetic fields \citep{Brun2005, Featherstone2009, Ratnasingam2024}. \citet{Barrault2025} provide an analytical study of inertial modes in the presence of a near-core toroidal field and investigate both uniform and differential rotation in their models. They find that the strength of the field is larger in the former case to reproduce the results of \citet{Saio2021}. These aspects directly impact the region of interest for our calculations and while they are not currently considered, investigating their influence in the future is highly relevant.

It follows from our discussion that a wide variety of field strengths are found both in observations and MHD simulations. Further work is required to link observations to simulations and shed light on possible observable conditions that can induce stronger fields. Valuable insight can be gained from characterizing a larger sample of main sequence intermediate-mass stars. This will allow us to compare with estimates of red giants and improve our understanding of core dynamo fields. 

\section{Conclusion}\label{sec:Conclusions}
In this work, we set upper limits on the near-core critical magnetic field strengths for three intermediate-mass stars, looking at both mixed modes and high-order g-modes. We determined the best MESA model for each star based on the stellar parameters, extracted frequencies with GYRE for the selected MESA profile, and determined the critical magnetic field strength with Dedalus, considering different field configurations during this process. Mode suppression is primarily driven by the radial component of the magnetic field, and our results show that the toroidal component does not significantly alter the critical field strength. We show that the mode geometry and magnetic field configuration influence the inferred magnetic field strength and that weaker fields can suppress $g$-modes in a quadrupole or mixed configuration. Although our results span a large range of magnetic field strengths, comparisons with observations of red giants and MHD simulations show that our estimates remain consistent with a core dynamo field, assuming that these can be strengthened by a fossil field in the case of KIC~3127996. We find evidence for rotational modulation in the two $\gamma$ Doradus stars, suggesting that they may host surface fields. We show that both stars experience little differential rotation, which is compatible with improved angular momentum transport due to magnetic fields. Our work provides insight into internal magnetic fields in $\gamma$-Doradus stars and is a necessary step towards characterizing magnetic fields and angular momentum transport in intermediate-mass stars on a larger scale. Future work on the inclusion of additional free parameters in the forward modeling and a more complete treatment of magnetic fields, for example with the incorporation of obliquity, will improve our capabilities for studying internal magnetic fields and mode suppression.

\begin{acknowledgements}
    The authors wish to thank May Gade Pedersen and Nuno Alexandre Martins Moedas for their valuable input in discussions regarding the MESA models. 
    This work is funded by the European Union (ERC, MAGNIFY, Project 101126182). However, the views and opinions expressed are those of the authors alone and do not necessarily reflect those of the European Union or the European Research Council. Neither the European Union nor the granting authority can be held responsible for them.
    DL is partially supported by NSF AAG grant AST-2405812, Sloan Foundation grant FG-2024-21548 and Simons Foundation grant SFI-MPS-T-MPS-00007353. ZG is supported by STFC grant UKRI1179.
\end{acknowledgements}

\bibliographystyle{aa} 
\bibliography{citations}

\begin{appendix}
    \section{MESA inlist}\label{sec:MESAinlist}
The typical MESA inlist for the main sequence evolution of an intermediate-mass star is shown below. \\

\noindent\&\texttt{star\_job} \par
   \texttt{load\_model\_filename = 'PMS.mod'}\par
   \texttt{save\_model\_filename = 'MS.mod'}\par
   \texttt{show\_log\_description\_at\_start = .true.}\par
   \texttt{load\_saved\_model = .true.}\par
   \texttt{history\_columns\_file = 'history\_columns.list'}\par
   \texttt{profile\_columns\_file = 'profile\_columns.list'}\par
   \texttt{pgstar\_flag = .false.}\par
   \texttt{change\_net = .true.}\par
   \texttt{change\_initial\_net = .true.}\par
   \texttt{new\_net\_name = 'pp\_cno\_extras\_o18\_ne22.net'}\par
   \texttt{save\_model\_when\_terminate = .true.} \\
/ \\

\noindent\&\texttt{kap}\par
   \texttt{use\_Type2\_opacities = .true.}\par
   \texttt{Zbase = } \\
/ \\

\noindent\&\texttt{controls} \par
   \texttt{max\_allowed\_nz = 60000}\par
   \texttt{mesh\_delta\_coeff = 0.4}\par
   \texttt{max\_dq = 1d-3}\par
   \texttt{varcontrol\_target = 1d-3} \par
   \texttt{timestep\_dt\_factor = 0.9} \par
   \texttt{energy\_eqn\_option = 'dedt'}\par
   \texttt{use\_gold2\_tolerances = .true.} \par
   \texttt{set\_min\_D\_mix = .true.}\par
   \texttt{min\_D\_mix = 10.0}\par
   \texttt{MLT\_option = 'Cox'}\par
   \texttt{mixing\_length\_alpha = 1.8}\par
   \texttt{use\_Ledoux\_criterion = .true.}\par
   \texttt{alpha\_semiconvection = 0d0}\par
   \texttt{thermohaline\_coeff = 0d0}\par
   \texttt{predictive\_mix(1) = .true.}\par
   \texttt{predictive\_zone\_type(1) = 'burn\_H'}\par
   \texttt{predictive\_zone\_loc(1) = 'core'}\par
   \texttt{predictive\_bdy\_loc(1) = 'top'}\par
   \texttt{overshoot\_scheme(1) = 'exponential'}\par
   \texttt{overshoot\_zone\_type(1) = 'burn\_H'}\par
   \texttt{overshoot\_zone\_loc(1) = 'core'}\par
   \texttt{overshoot\_bdy\_loc(1) = 'top'}\par
   \texttt{overshoot\_f(1) = 0.017}\par
   \texttt{overshoot\_f0(1) = 0.002}\par
   \texttt{overshoot\_D\_min = 1d-2}\par
   \texttt{overshoot\_brunt\_B\_max = 0d0}\par
   \texttt{num\_cells\_for\_smooth\_brunt\_B = 0}\par
   \texttt{num\_cells\_for\_smooth\_gradL\_composition\_term=0}\par
   \texttt{remove\_mixing\_glitches = .false.}\par
   \texttt{write\_pulse\_data\_with\_profile = .true.}\par
   \texttt{pulse\_data\_format = 'GYRE'}\par
   \texttt{add\_atmosphere\_to\_pulse\_data = .false.}\par
   \texttt{keep\_surface\_point\_for\_pulse\_data = .true.}\par
   \texttt{add\_double\_points\_to\_pulse\_data = .true.}\par
   \texttt{threshold\_grad\_mu\_for\_double\_point = 5d0}\par
   \texttt{interpolate\_rho\_for\_pulse\_data = .true.}\par
   \texttt{log\_center\_density\_upper\_limit = 3.25} \par
   \texttt{max\_num\_profile\_models = 500}\par
   \texttt{photo\_interval = 500} \\
/ 

\section{GYRE inlist}\label{sec:GYREinlist}
An example GYRE inlist is shown below, that extracts frequencies of orders $n_g=-100,-15$ of the $(\ell, m)=(1,1)$ mode for a given MESA profile number XX. The rotation frequency is set to 0 in this example but is modified according to the inferred rotation rate of the target star.  \\ \\
\&\texttt{model} \par
   \texttt{model\_type = 'EVOL'} \par
   \texttt{file = 'profileXX.data.GYRE'} \par
   \texttt{file\_format = 'MESA'} \\
/ 

\noindent\&\texttt{mode} \par

   \texttt{l = 1} \par
   \texttt{m = 1} \par
   \texttt{n\_pg\_min = -100} \par
   \texttt{n\_pg\_max = -15} \\
/

\noindent \&\texttt{osc} \par
   \texttt{outer\_bound = 'VACUUM'} \\
/

\noindent \&\texttt{rot} \par
   \texttt{Omega\_rot\_source = 'UNIFORM'} \par
   \texttt{Omega\_rot = 0.0} \par
   \texttt{Omega\_rot\_units = 'CYC\_PER\_DAY'} \par
   \texttt{coriolis\_method = 'TAR'} \\
/

\noindent \&\texttt{num} \par
   \texttt{diff\_scheme = 'COLLOC\_GL4'} \par
   \texttt{n\_iter\_max = 30} \\
/

\noindent \&\texttt{scan} \par
   \texttt{grid\_type = 'INVERSE'} \par
   \texttt{grid\_frame = 'INERTIAL'} \par
   \texttt{freq\_units = 'CYC\_PER\_DAY'} \par
   \texttt{freq\_min = 0.3} \par
   \texttt{freq\_max = 5} \par
   \texttt{n\_freq = 1000} \\
/

\noindent \&\texttt{grid} \par
   \texttt{w\_osc = 10} \par
   \texttt{w\_ctr = 10} \par
   \texttt{w\_exp = 2} \\
/

\noindent \&\texttt{ad\_output} \par
   \texttt{freq\_units = 'CYC\_PER\_DAY'} \par
   \texttt{summary\_file = 'profileXX\_l1\_m1-freqs.dat'} \par
   \texttt{summary\_file\_format = 'TXT'} \par
   \texttt{summary\_item\_list = 'l, m, n\_pg, n\_p, n\_g,} \par \texttt{freq, omega, E\_norm'} \par
   \texttt{detail\_template = 'profileXX\_l\%l\%m\%n.txt'} \par
   \texttt{detail\_file\_format = 'TXT'} \par
   \texttt{detail\_item\_list = 'Gamma\_1, V\_2, c\_1, x,} \par \texttt{xi\_r, xi\_h, Delta\_p, Delta\_g, H, W\_eps,} \par \texttt{dW\_dx, dE\_dx, nabla\_ad, eta, rho'} \\
/

\section{Observed frequencies}\label{sec:Frequencies}
\begin{table}[htp]
\centering
\caption{Observed modes for KIC~3127996.}
\begin{tabular}{cc|cc|cc}
    \hline
     \multicolumn{2}{c|}{$(\ell, m)=(1,1)$} & \multicolumn{2}{c|}{$(\ell, m)=(1,0)$} & \multicolumn{2}{c}{$(\ell, m)=(1,-1)$} \\
     $n_{pg}$ & $f_{\rm{obs}}\, [\rm{d}^{-1}]$ & $n_{pg}$ & $f_{\rm{obs}}\, [\rm{d}^{-1}]$ & $n_{pg}$ & $f_{\rm{obs}}\, [\rm{d}^{-1}]$ \\
     \hline
     $-31$ & 0.8957 & $-30$ & 0.9232 & $-31$ & 0.8687 \\
     $-30$ & 0.9232 & $-29$ & 0.9503 & $-30$ & 0.8957 \\
     $-29$ & 0.9520 & $-28$ & 0.9791 & $-29$ & 0.9250 \\
     $-28$ & 0.9827 & $-27$ & 1.0098 & $-28$ & 0.9556 \\
     $-27$ & 1.0151 & $-25$ & 1.0768 & $-27$ & 0.9881 \\
     $-26$ & 1.0496 & $-24$ & 1.1139 & $-26$ & 1.0226 \\
     $-25$ & 1.0868 & $-23$ & 1.1538 & $-25$ & 1.0596 \\
     $-24$ & 1.1266 & $-22$ & 1.1971 & $-24$ & 1.0995 \\
     $-22$ & 1.2169 & $-21$ & 1.2441 & $-23$ & 1.1427 \\
     $-21$ & 1.2679 & & & $-22$ & 1.1898 \\
     $-20$ & 1.3230 & & & & \\
     \hline
     \multicolumn{2}{c|}{$(\ell, m)=(2,1)$} & \multicolumn{2}{c|}{$(\ell, m)=(2,-2)$} & \multicolumn{2}{c}{} \\
     $n_{pg}$ & $f_{\rm{obs}}\, [\rm{d}^{-1}]$ & $n_{pg}$ & $f_{\rm{obs}}\, [\rm{d}^{-1}]$ & & \\
     \hline
     $-35$ & 1.3950 & $-35$ & 1.3053 & & \\
     $-34$ & 1.4305 & $-34$ & 1.3406 & & \\
     $-33$ & 1.4679 & $-33$ & 1.3781 & & \\
     $-32$ & 1.5076 & $-32$ & 1.4594 & & \\
     $-31$ & 1.5492 & $-31$ & 1.5048 & & \\
     $-30$ & 1.5935 & $-30$ & 1.5546 & & \\
     \hline
     \end{tabular}
     \tablefoot{The table is divided in six columns, corresponding to each $(\ell,m)$ pair observed for the star. We report the order $n_{\rm{pg}}$ and frequency $f_{\rm{obs}}$ for each mode, according to the procedure detailed in Section \ref{sec:FreqExtract}.}
     \label{tab:KIC3217996}
\end{table}

\begin{table}[htp]
    \centering
    \caption{Observed modes in 44 Tau.}
    \begin{tabular}{cccc}
        \hline 
        $\ell$ & $n$ & $f_{\rm{obs}}\ [\rm{d^{-1}}]$ & $\%_{\rm{g}}$ \\
        \hline
        2 & $g_{10}$ & 5.3047 & 96 \\
        2 & $g_8$ & 6.3390 & 83 \\
        2 & $g_7$ & 6.7955 & 68 \\
        2 & $g_5$ & 9.5828 & 62 \\
        2 & $g_4$ & 11.2947 & 60\\
        2 & $p_1$ & 8.6391 & 64 \\
        2 & $p_2$ & 12.6115 & 34 \\
        \hline
    \end{tabular}
    \tablefoot{$n$ shows the order of the dominant component of the mixed mode, with $g$ ($p$) indicating it is a primarily $g$-mode (p-mode) dominated mode. The last column indicates the fraction of the mode that is concentrated in the near-core region. The values are taken from \citet{Lenz2010}.}
    \label{tab:44TauSummary}
\end{table}

\begin{table}[htp]
\caption{Observed modes for KIC~5876187.}
    \begin{tabular}{cc|cc|cc}
        \hline
         \multicolumn{2}{c|}{$(\ell, m)=(1,1)$} & \multicolumn{2}{c|}{$(\ell, m)=(2,2)$} & \multicolumn{2}{c}{$(\ell, m)=(3,3)$} \\
         $n_{pg}$ & $f_{\rm{obs}}\, [\rm{d}^{-1}]$ & $n_{pg}$ & $f_{\rm{obs}}\, [\rm{d}^{-1}]$ & $n_{pg}$ & $f_{\rm{obs}}\, [\rm{d}^{-1}]$ \\
         \hline
         $-111$ & 0.7814 & $-92$ & 1.6515 & $-86$ & 2.5237 \\
         $-109$ & 0.7885 & $-89$ & 1.6672 & $-85$ & 2.5323 \\
         $-107$ & 0.7922 & $-87$ & 1.6784 & $-83$ & 2.5503 \\
         $-105$ & 0.7961 & $-86$ & 1.6841 & $-77$ & 2.6102 \\
         $-102$ & 0.7981 & $-83$ & 1.7023 & $-76$ & 2.6211 \\
         $-101$ & 0.8022 & $-82$ & 1.7087 & $-75$ & 2.6323 \\
         $-98$ & 0.8044 & $-81$ & 1.7152 & $-73$ & 2.6556 \\
         $-97$ & 0.8111 & $-79$ & 1.7287 & $-72$ & 2.6677 \\
         $-95$ & 0.8134 & $-78$ & 1.7358 & $-70$ & 2.6930 \\
         $-94$ & 0.8182 & $-76$ & 1.7504 & $-67$ & 2.7337 \\
         $-93$ & 0.8207 & $-75$ & 1.7580 & $-65$ & 2.7779 \\
         $-92$ & 0.8233 & $-74$ & 1.7658 & $-63$ & 2.8096 \\
         $-90$ & 0.8259 & $-73$ & 1.7739 & & \\
         $-89$ & 0.8312 & $-71$ & 1.7905 & & \\
         $-87$ & 0.8340 & $-68$ & 1.8174 & & \\
         $-86$ & 0.8369 & $-65$ & 1.8465 & & \\
         $-85$ & 0.8398 & $-63$ & 1.8673 & & \\
         $-84$ & 0.8428 & $-62$ & 1.8782 & & \\
         $-83$ & 0.8458 & $-61$ & 1.9010 & & \\
         $-82$ & 0.8490 & $-60$ & 1.9129 & & \\
         $-81$ & 0.8522 & $-54$ & 1.9924 & & \\
         $-80$ & 0.8555 & & & & \\
         $-78$ & 0.8588 & & & & \\
         $-77$ & 0.8622 & & & & \\
         $-75$ & 0.8694 & & & & \\
         $-74$ & 0.8731 & & & & \\
         $-72$ & 0.8808 & & & & \\
         $-71$ & 0.8847 & & & & \\
         $-69$ & 0.8930 & & & & \\
         $-68$ & 0.8973 & & & & \\
         $-67$ & 0.9063 & & & & \\
         $-66$ & 0.9109 & & & & \\
         $-65$ & 0.9157 & & & & \\
         $-64$ & 0.9206 & & & & \\
         $-63$ & 0.9256 & & & & \\
         $-62$ & 0.9308 & & & & \\
         $-60$ & 0.9361 & & & & \\
         $-59$ & 0.9416 & & & & \\
         $-56$ & 0.9589 & & & & \\
         $-55$ & 0.9650 & & & & \\
         \hline
    \end{tabular}
    \tablefoot{The layout is identical to that presented in Table \ref{tab:KIC3217996}.}
    \label{tab:KIC5876187}
\end{table}
\newpage
\section{Dedalus equations}\label{sec:DedalusEq}
We provide in this section the equations used in Dedalus to solve the eigenvalue problem and determine the critical magnetic field strengths. For simplicity, we write the magnetic field components as $B_r$ and $B_\phi$ for the poloidal and toroidal fields, respectively. These include the spherical harmonics, as shown in Eqs. \ref{eq:Poloidal} and \ref{eq:Toroidal}. The horizontal momentum equation, Eq. \ref{eq:MagneticField}, can thus be written: 
\begin{equation}
     -i\omega\boldsymbol{u}_h+2\boldsymbol{\Omega}\times\boldsymbol{u}_h+\frac{1}{r}\boldsymbol{\nabla}_hp=\left(\frac{ik_rB_r}{4\pi\rho_0}+\frac{B_\phi}{4\pi\rho_0r}\boldsymbol{e}_\phi\boldsymbol{\cdot}\boldsymbol{\nabla}_h\right)\boldsymbol{b}_h\,.
     \label{eq:SimpleHM}
\end{equation}
We can eliminate the magnetic field perturbation, $\boldsymbol{b}_h$, from this equation using the induction equation. This reduces the number of parameters in our problem. Isolating for the magnetic field in Eq. \ref{eq:Lecoanet2} yields:  
\begin{equation}
    \boldsymbol{b}_h=-\frac{1}{i\omega}\left(iB_rk_r\boldsymbol{u}_h+B_\phi\boldsymbol{e}_\phi\boldsymbol{\cdot}\boldsymbol{\nabla}_h\boldsymbol{u}_h\right)\,.
\end{equation}
The Lorentz force, on the right-hand side of Eq. \ref{eq:SimpleHM}, can thus be written: 
\begin{equation}
    -\left(\frac{ik_rB_r}{4\pi\rho_0}+\frac{B_\phi}{4\pi\rho_0}\boldsymbol{e}_\phi\boldsymbol{\cdot}\boldsymbol{\nabla}_h\right)\frac{1}{i\omega}(iB_rk_r\boldsymbol{u}_h+B_\phi\boldsymbol{e}_\phi\boldsymbol{\cdot}\boldsymbol{\nabla}_h\boldsymbol{u}_h)\,,
\end{equation}
Which yields three terms when expanding:
\begin{equation}
    \frac{1}{4\pi\omega\rho_0}\left(\frac{k_r^2}{i}B_r^2-\frac{1}{ir^2}B_\phi\boldsymbol{e}_\phi\boldsymbol{\cdot}\boldsymbol{\nabla}_hB_\phi\boldsymbol{e}_\phi\boldsymbol{\cdot}\boldsymbol{\nabla}_h-\frac{2k_r}{r}B_rB_\phi\boldsymbol{e}_\phi\boldsymbol{\cdot}\boldsymbol{\nabla}_h\right)\boldsymbol{u}_h\,.
\end{equation}
Inserting in Eq. \ref{eq:SimpleHM} provides the final expression, after rearranging some terms: 
\begin{equation}\begin{split}
\omega\boldsymbol{u}_h+2i\boldsymbol{\Omega}\times\boldsymbol{u}_h&+\frac{i}{r}\boldsymbol{\nabla}_hp  -\frac{k_r^2B_r^2}{4\pi\omega\rho_0}\boldsymbol{u}_h\\
& +\frac{1}{4\pi\omega\rho_0r^2}B_\phi\boldsymbol{e}_\phi\boldsymbol{\cdot}\boldsymbol{\nabla}_hB_\phi\boldsymbol{e}_\phi\boldsymbol{\cdot}\boldsymbol{\nabla}_h\boldsymbol{u}_h \\
& + \frac{2ik_r}{2\pi\omega\rho_0r}B_rB_\phi\boldsymbol{e}_\phi\boldsymbol{\cdot}\boldsymbol{\nabla}_h\boldsymbol{u}_h=0\,.
\end{split}
\label{eq:DedalusHM}
\end{equation}
Eq. \ref{eq:DedalusHM}, along with the continuity and radial momentum equations, Eqs. \ref{eq:Continuity}-\ref{eq:Radial}, constitute the full set of equations. 

\end{appendix}

\end{document}